\RequirePackage{amsmath}

\documentclass[review]{elsarticle}

\usepackage{graphicx}
\usepackage{caption}
\usepackage{subcaption}
\usepackage{mathtools}
\usepackage{multirow}
\usepackage{algorithm}
\usepackage[noend]{algpseudocode}
\usepackage[utf8x]{inputenc}
\usepackage[dvipsnames]{xcolor}
\usepackage[hidelinks=true, pdffitwindow=true, setpagesize=false]{hyperref}
\usepackage{listings}
\usepackage{color}

\lstdefinestyle{mystyle}{
	backgroundcolor=\color{white},   
	commentstyle=\color{green},
	keywordstyle=\color{orange},
	numberstyle=\tiny\color{black},
	stringstyle=\color{red},
    basicstyle=\small,
	breakatwhitespace=false,         
	breaklines=true,                 
	captionpos=b,                    
	keepspaces=true,                 
	numbers=none,
	numbersep=5pt,                  
	showspaces=false,                
	showstringspaces=false,
	showtabs=false,                  
	tabsize=2,
	frame=none    
}

\usepackage{ifsym}

\begin{document}
	\title{Benchmarking Top-K Keyword and Top-K Document Processing with T$^2$K$^2$ and T$^2$K$^2$D$^2$}
    
\begin{frontmatter}
    \author{Ciprian-Octavian Truică$^{1,a}$,
    J{\'e}r{\^o}me Darmont$^{2,b}$, \\
    Alexandru Boicea$^{1,c}$, 
    Florin Rădulescu$^{1,d}$}
	\address{$^{1}$Computer Science and Engineering Department, Faculty of Automatic Control and Computers, University Politehnica of Bucharest, Bucharest, Romania\\
    $^{2}$Universit\'{e} de Lyon, Lyon 2, ERIC EA 3083, France\\
$^{a}$ciprian.truica@cs.pub.ro,  
$^{b}$jerome.darmont@univ-lyon2.fr,\\
$^{c}$alexandru.boicea@cs.pub.ro,
$^{d}$florin.radulescu@cs.pub.ro
}

\begin{abstract}
	Top-$k$ keyword and top-$k$ document extraction are very popular text analysis techniques. Top-$k$ keywords and documents are often computed on-the-fly, but they exploit weighted vocabularies that are costly to build. To compare competing weighting schemes and database implementations, benchmarking is customary. To the best of our knowledge, no benchmark currently addresses these problems. Hence, in this paper, we present T$^2$K$^2$, a top-$k$ keywords and documents benchmark, and its decision support-oriented evolution T$^2$K$^2$D$^2$. Both benchmarks feature a real tweet dataset and queries with various complexities and selectivities. They help evaluate weighting schemes and database implementations in terms of computing performance. To illustrate our benchmarks' relevance and genericity, we successfully ran performance tests on the TF-IDF and Okapi BM25 weighting schemes, on one hand, and on different relational (Oracle, PostgreSQL) and document-oriented (MongoDB) database implementations, on the other hand.
\end{abstract}
    
    
\end{frontmatter}

\section{Introduction}
\label{sec:Intro}

	With the tremendous and continuous development of textual web contents, especially (but not only) on social media, text analytics has become a pivotal trend that still proposes many challenges nowadays. Among text analytics techniques, top-$k$ keyword and document extraction is very popular. For instance, extracting the $k$ most frequent terms from a corpus helps determine trends \cite{Ravat2008,Bringay2011} or detect specific events \cite{Guille2015}; and finding the $k$ documents that are the most similar to a query is of course the core task of search engines. Moreover, computing top-$k$ keywords and documents requires building a weighted vocabulary, which can also be used for many other text analytics purposes such as topic modeling~\cite{AlghamdiK2015} and clustering \cite{AggarwalZ12}. 
	
	To compare combinations of weighting schemes, computing strategies and physical implementations, benchmarking is customary. However, most prominent big data benchmarks \cite{Huang2010,Wang2014,tpcxhs15} focus on MapReduce operations and do not specifically model text-oriented workflows. Very few benchmarks do, but still focus on Hadoop-like architectures and remain at the methodology \cite{Gattiker2013} and specification \cite{Ferrarons2014} stage. Moreover, computing weighting schemes at the application level can prove inefficient when working with large data volumes, because all the information must be queried, read and processed at a different layer but storage. A better approach is to process the information at the storage layer using aggregation functions and then return the response to the application layer. 

	Hence, we introduced the Twitter Top-K Keywords Benchmark (T$^2$K$^2$)~\cite{Truica2017}, which relies on database storage and features a real tweet dataset, as well as queries with various complexities and selectivities. We also designed T$^2$K$^2$ to be somewhat generic, i.e., it can compare various weighting schemes, database logical and physical implementations and even text analytics platforms in terms of computing efficiency. In this paper, we further this work with the following contributions.
	\begin{enumerate}
		\item Since T$^2$K$^2$'s data model is generic enough to handle any type of textual documents (not only tweets), we complement its workload model with top-$k$ document queries.
		\item We illustrate T$^2$K$^2$'s relevance and genericity by performing tests on the TF-IDF and Okapi BM25 weighting schemes, on one hand, and on different relational (Oracle, PostgreSQL) and document-oriented (MongoDB) database implementations, on the other hand.
		\item Since T$^2$K$^2$ queries are analytical by nature, we hypothesize that a star (data warehouse) schema should improve query response and propose an evolution of T$^2$K$^2$ named T$^2$K$^2$D$^2$, where the last two Ds stand for Dimensional and Documents, respectively.
		\item We complement our first experiments with top-$k$ keyword and top-$k$ document benchmarking with T$^2$K$^2$D$^2$, still testing the TF-IDF and Okapi BM25 weighting schemes, as well as the Oracle, PostgreSQL and MongoDB database systems.
	\end{enumerate}
	
	We designed our benchmarks' queries by logging computational linguists working on a text analysis platform~\cite{Truica2016a,Truica2016b} on real-world data. After analyzing and clustering similar user queries, we ended up with 8 queries (4 for T$^2$K$^2$ and 4 for T$^2$K$^2$D$^2$) that we consider generic enough to benchmark any similar system.

	Moreover, note that our benchmarks do not apply to Information Retrieval (IR) system evaluation. They rather target database systems built on top of an IR system, with the aim of facilitating text mining, e.g., text classification, topic modeling or event detection. Benchmark queries are relevant to machine learning tasks, where it is important to analyze different subsets of the dataset to extract knowledge. In contrast, IR systems do not handle subsets of the initial corpus well, nor datasets whose size changes in time, because they compute weights only once. However, this creates two result reproducibility problems~\cite{Lin2016}. First, when using subsets of the initial corpus, weights are not recomputed, which induces errors when computing the ranking functions for top-$k$ keywords and documents. Second, weights must be recomputed each time the size of the dataset changes, inducing more write operations that are computationally demanding.
	
	The database approach can solve these result reproducibility problems by dynamically computing weights at search time using fielded data. Although some IR systems do use fielded data~\cite{Crane2017}, as they systems compute weights only once, ranking functions do not update weights when subsets of the initial corpus are used or when the volume of data changes~\cite{Bellot2013}, which induces errors during the extraction of top-$k$ documents and keywords. Moreover, computing weights dynamically is ideal for analyzing subsets of the initial dataset or corpora whose size changes over time, e.g., when analyzing stream data~\cite{Bifet2010}. Database systems are designed to manage large volumes of data with a high throughput and can easily manipulate changing volumes of data by using optimized CRUD (Create, Read, Update, Delete) operations. Thus, they are better suited for handling datasets whose size changes over time and analyzing subsets of the initial corpus.
	
	The remainder of this paper is organized as follows. In Section~\ref{sec:RelatedWorks}, we survey existing big data and more specifically text processing-oriented benchmarks. In Section~\ref{sec:Specs}, we recall T$^2$K$^2$'s data and workload models, and provide T$^2$K$^2$D$^2$'s full specifications. In Section~\ref{sec:Implementations}, we detail how we implement two weighting schemes (namely, TF-IDF and Okapi BM25) and instantiate our benchmarks in Oracle, PostgreSQL and MongoDB. In Section~\ref{sec:Experiments}, we account for the experiments we performed with both T$^2$K$^2$ and T$^2$K$^2$D$^2$, which demonstrate the feasibility and relevance of our benchmarks. Finally, we conclude the paper and provide research perspectives in Section~\ref{sec:Conclusion}.

\section{Related Works}
\label{sec:RelatedWorks}

\subsection{Big Data Benchmarks}

	There exist many big data benchmarks, which are mostly data-centric, i.e., they focus on volume and MapReduce-based applications, rather than on variety, and do not include textual data. 

	For instance, the quasi-standard TPCxHS benchmark models a simple application and features, in addition to classical throughput and response time metrics, availability and energy metrics \cite{tpcxhs15}. Similarly, HiBench is a set of Hadoop programs, ranging from data sorting to clustering, aimed at measuring metrics such as response time, HDFS bandwidth consumption and data access patterns~\cite{Huang2010}.

	In contrast, BigDataBench, among 19 benchmarks, features application scenarios from search engines, i.e., the application on Wikipedia entries of operators such as Grep or WordCount \cite{Wang2014}. Yet, although BigDataBench is open source, it is quite complex and difficult to extend to test the computation efficiency of term weighting schemes.

\subsection{Text Analysis Benchmarks}

	Term weighting schemes are extensively benchmarked in several subdomains of text analytics. In sentiment analysis, dictionary-based methods are benchmarked on various types of corpora (news articles, movie reviews, books and tweets) \cite{Reagan2015} to perform both quantitative and qualitative assessments. 

	In text classification and categorization, benchmarks target different types of texts: short texts represented by sentence pairs with similarity ratings \cite{OShea2010}; large texts from Reuters newswire stories on one hand \cite{Lewis2004}, and DBpedia and the Open Directory Project on the other hand \cite{Partalas2015}; or texts in a specific language~\cite{Kilinc2017}.

	In terms of metrics, all the above-mentioned benchmarks focus on algorithm accuracy. Either term weights are known before the algorithm is applied, or their computation is incorporated with preprocessing. Thus, such benchmarks do not evaluate weighting scheme construction efficiency as we do.

	Finally, TextGen is a synthetic textual data generator \cite{Wang2016}. TextGen builds corpora by segmenting real-world text datasets and enforces a lognormal word-frequency distribution. However, TextGen aims at testing the compression performance of word-based compressors, and thus provides no workload or metrics that are suitable to our needs.

\subsection{Parallel Text Processing Benchmarks}

	This last family of benchmarks evaluates parallel text processing in big data, cloud applications. However, it is very small (two benchmarks only).
The first one is actually a methodology for designing text-oriented benchmarks in Hadoop~\cite{Gattiker2013}. It provides both guidelines and solutions for data preparation and workload definition. Yet, as text analysis benchmarks, its metrics measure the accuracy of analytics results, while we aim at evaluating computing performance.

	PRIMEBALL features a fictitious news site hosted in the cloud that is to be managed by the framework under analysis, together with several objective use cases and measures for evaluating system performance \cite{Ferrarons2014}. One of its metrics notably involves searching a single word in the corpus. However, PRIMEBALL remains a specification only as of today.

\section{Benchmark Specifications}
\label{sec:Specs}

	Text analysis deals with discovering hidden patterns from texts. In most cases, it is useful to determine such patterns for given groups, e.g., males and females, because they have different interests and talk about disjunct subjects. Moreover, if new events appear, depending on the location and time of day, these subject can change for the same group of people. The queries we propose aim to determine such hidden patterns and improve text analysis and anomaly detection.
		
	Typically, a benchmark  is  constituted  of a  data  model  (conceptual  schema  and extension), a  workload  model (set  of operations) to  apply  on  the dataset, an execution protocol and performance  metrics~\cite{Darmont2017}. In this section, we provide a conceptual description of T$^2$K$^2$ and T$^2$K$^2$D$^2$, so that it is generic and can cope with various weighting schemes and database logical and physical implementations. For generalization purposes, the models will contain the words "document" and "documents" for naming the entities, instead of words "tweet" and "tweets".

\subsection{Text Preprocessing}
	
	Text preprocessing is an important task used in Text Mining and Analysis, Information Retrieval and Natural Language processing. Text preprocessing involves cleaning and preparing texts for different tasks, e.g., classification, IR or sentiment analysis. Usually, textual data contain lot of noise and uninformative parts such as HTML tags, scripts and links. At the word level, common words impact negatively the task at hand.  Removing these so-called stop words decreases the problem's dimensionality~\cite{Haddi2013}. Moreover, preprocessing is used to extract information and metadata from text, standardize the document corpus, remove useless information and minimize the dimensionality of the data. 

	In our benchmarks, tweets are preprocessed \emph{a priori}. Lemmas are extracted from each word and stop words removed to minimize vocabulary size and improve the accuracy of machine learning algorithms such as topic modeling and document clustering. End users can preprocess data as they see fit and then store the information in the database using the models provided by the benchmark. In our preprocessing step, words that are abbreviated are considered words in their own sense, i.e., they are not expanded. For instance, we consider "omg" as a word on his own. Misspellings are currently retained, although in a next version of the preprocessing step, we plan to address this problem. Moreover, abbreviations and misspellings only affect the selectivity of queries, as the vocabulary stores them too.

	The preprocessing steps we apply on raw texts to construct the weighted vocabulary follow.
    \begin{enumerate}
    	\item For each tweet, we extract tags, i.e., hashtags (which make good keywords), attags, and remove links.
	    \item We expand contractions, i.e., shortened versions of the written and spoken forms of a word, syllable, or word group, created by omission of internal letters and sounds~\cite{Cooper2014}. For example, "it's" becomes "it is".
    	\item We split the text of a tweet into sentences.
	    \item We extract the part of speech for each sentence.
	    \item We build a clean text by removing punctuation and stop words.
	    \item For each term in a clean text, we use the part of speech to extract lemmas and create a lemma text.
    	\item For each lemma $t$ in lemma text $d$, we compute the number of co-occurrences $f_{t,d}$ and term frequency $TF(t,d)$, which normalizes $f_{t,d}$.
    \end{enumerate}

\subsection{T$^2$K$^2$}
\label{sec:SpecT2K2}

\subsubsection{Data Model}
\label{sec:T2K2data}

\begin{figure}[!htbp]
        \begin{center}
            \includegraphics[width=\columnwidth]{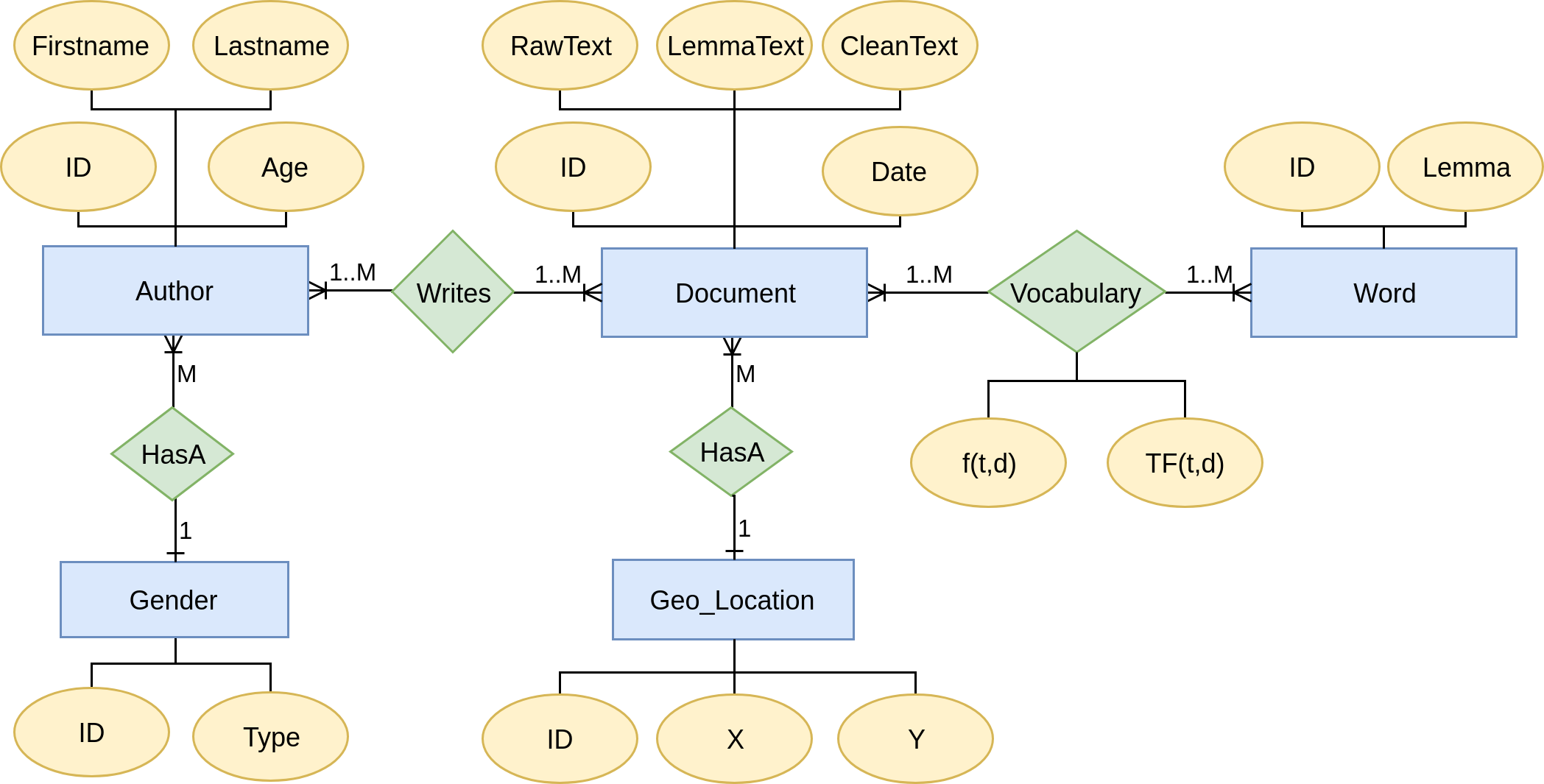}
            \caption{T$^2$K$^2$ conceptual data model} 
            \label{fig:T2K2ERDiagram}
        \end{center}
    \end{figure}

	All the information and metadata extracted during the preprocessing step are modeled using the entity-relationship presented in Figure~\ref{fig:T2K2ERDiagram}. Let us describe all entities in this conceptual model.
\begin{itemize}
    \item \textit{Author} stores information about a tweet author, e.g., unique identifier, firstname, lastname, and age.
    \item \textit{Gender} is used to store the authors gender and to remove duplication in the database.
    \item \textit{Document} entity stores all the textual information and metadata for a tweet, e.g., the unique identifier of the tweet, the date, and the original and the processed text of the text, i.e., original text (\emph{RawText}), clean text (\emph{CleanText}) and lemma (\emph{LemmaText}).
    \item \emph{Writes} is the many-to-many relationship between an author and a tweet.
    \item to eliminate duplicates and minimize the storage space, the \emph{Geo\_Location} is used to store the geo-location for each tweet.
    \item \textit{Word} entity stores the lemma and its unique identifier.
    \item \emph{Vocabulary} is a many-to-many relationship between the \textit{Word} and \textit{Document} entities that stores, beside the unique identifiers of the lemma and tweet, the number of co-occurrence $f_{t,d}$ and term frequency $TF(t,d)$ for a lemma in a tweet.
\end{itemize}

\subsubsection{Workload Model}
\label{sec:T2K2workload}

	T$^2$K$^2$ features 4 queries, from $Q1$ to $Q4$, which are grouping queries that extract the top-$k$ keywords with different constraints that are combined in the queries: $c_1 (Q1)$, $c_1 \wedge c_2 (Q2)$, $c_1 \wedge c_3 (Q3)$ and $c_1 \wedge c_2 \wedge c_3 (Q4)$. We define constraints  $c_1$ to $c_3$ below.
\begin{itemize}
    \item $c_1$ is \emph{Gender.Type = pGender} with \emph{pGender $\in$ \{male, female\}}.
    \item $c_2$ is \emph{Document.Date $\in$ [pStartDate, pEndDate]}, where \emph{pStartDate, pEndDate $\in$ [2015-09-17 20:41:35, 2015-09-19 04:05:45]} and \emph{pStartDate $<$ pEndDate}.
    \item $c_3$ is \emph{Geo\_location.X $\in$ [ pStartX, pEndX]} and \emph{Geo\_Location.Y $\in$ [pStartY, pEndY]}, where \emph{pStartX, pEndX $\in$ [15, 50]}, \emph{pStartX $<$ pEndX}, \emph{pStartY, pEndY $\in$ [-124, 120]} and \emph{pStartY $<$ pEndY}.
\end{itemize}

	For extracting top-$k$ documents, an additional constraint $c_4$ on the terms of the vocabulary is needed. This constraint is used to select only the tweets that contain at least one of the word from a list of search terms and construct the tweets hierarchy only for them. Therefore, $c_4$ is \emph{Words.Lemma = pTerms} where \emph{pTerms $\in$ \{ term $\mid$ term $\in$ vocabulary \} }.  The queries in this case rewritten using the constraints $c_1$ to $c_4$ as $c_1 \wedge c_4 (Q1)$, $c_1 \wedge c_2 \wedge c_4 (Q2)$, $c_1 \wedge c_3 \wedge c_4 (Q3)$ and $c_1 \wedge c_2 \wedge c_3 \wedge c_4 (Q4)$.

\subsection{T$^2$K$^2$D$^2$}
\label{sec:SpecT2K2D2}

\subsubsection{Data Model}
\label{sec:T2K2D2data}

\begin{figure}[!htbp]
        \begin{center}
            \includegraphics[width=\columnwidth]{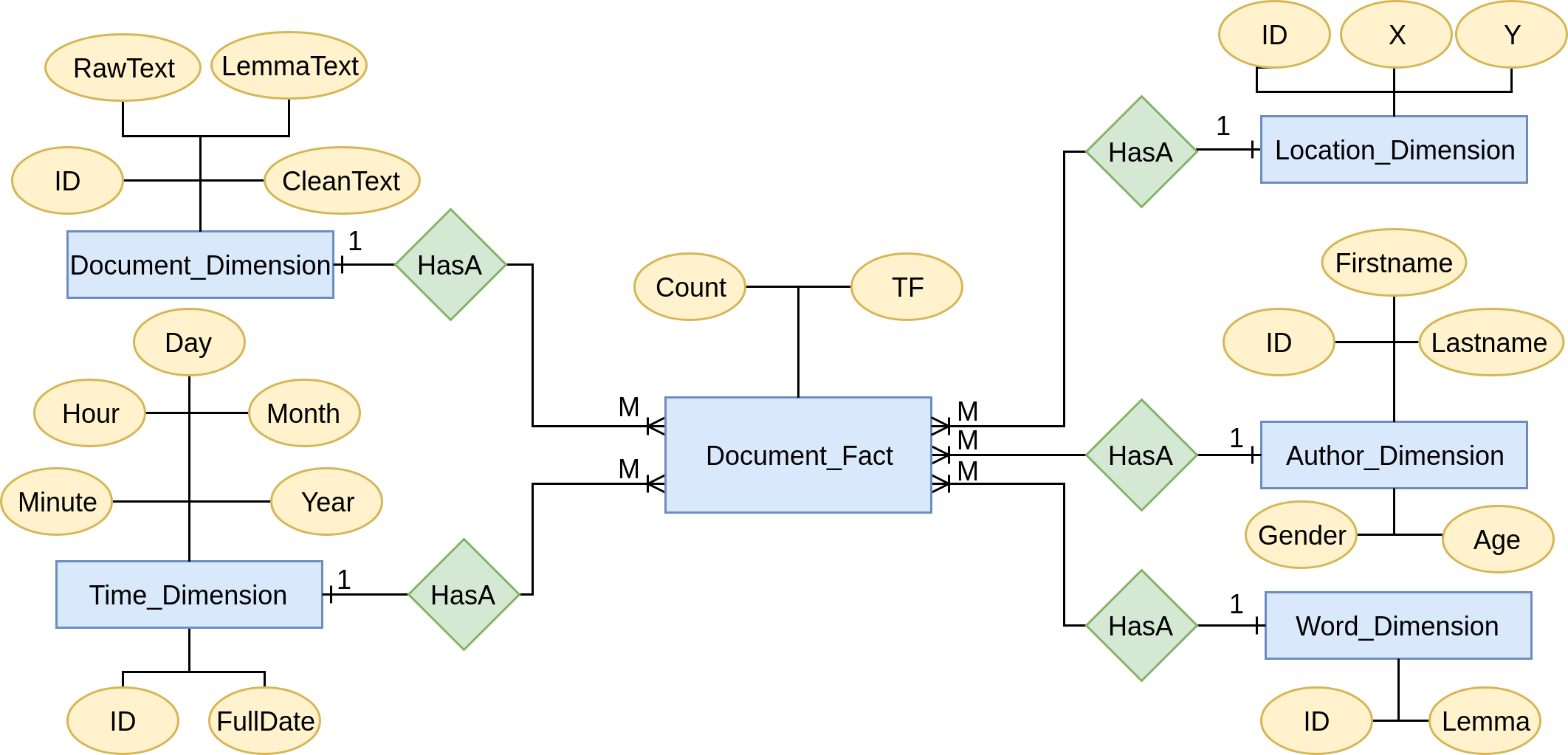}
            \caption{T$^2$K$^2$D$^2$ conceptual data model} 
            \label{fig:ERStarDiagram}
        \end{center}
\end{figure}

T$^2$K$^2$'s workload model (Section~\ref{sec:T2K2workload}) is made of grouping queries akin to OLAP queries. Thus, to improve query performance, we remodel its schema into a multidimensional star schema (Figure~\ref{fig:ERStarDiagram}) featuring a central fact table and multiple dimensions. T$^2$K$^2$D$^2$'s schema contains the following entities.
\begin{itemize}
    \item \emph{Document\_Fact} is the central fact entity and contains the number of co-occurrence $f_{t,d}$ and term frequency $TF(t,d)$ for a lemma in a document.
    \item \emph{Document\_Dimension} is the tweets dimension table, containing each tweet's unique identifier and the original and processed text, i.e., original text (\emph{RawTtext}), clean text (\emph{CleanText}) and lemma text(\emph{LemmaText}).
    \item \emph{Word\_Dimension} stores the lema and its unique identifier.
    \item \emph{Time\_Dimension} stores the full date and also its hierarchy composed of minute, hour, day, month and year 
    \item \emph{Author\_Dimension} is the author dimension and stores information about an author's unique identifier, gender, age, firstname, and lastname.
    \item \emph{Location\_Dimension} is the geo-location dimension. 
\end{itemize}

\subsubsection{Workload Model}
\label{sec:T2K2D2workload}

	To adjust to T$^2$K$^2$D$^2$'s multidimensional data model, we rewrite T$^2$K$^2$'s queries as OLAP queries. We retain the same constraints $c_1$ to $c_4$, only adapting entity and attribute names. The redefined constraints' specification follows.
\begin{itemize}
    \item $c_1$ is \emph{Autor\_Dimension.gender = pGender} with  \emph{pGender $\in$ \{male, female\}}.
    \item $c_2$ is \emph{Time\_Dimension.Date $\in$ [pStartDate, pEndDate]}, where \emph{pStartDate, pEndDate $\in$ [2015-09-17 20:41:35, 2015-09-19 04:05:45]} and \emph{pStartDate $<$ pEndDate}.
    \item $c_3$ is \emph{Location\_Dimension.X $\in$ [ pStartX, pEndX]} and \emph{Location\_Dimension.Y $\in$ [pStartY, pEndY]}, where \emph{pStartX, pEndX $\in$ [15, 50]}, \emph{pStartX $<$ pEndX}, \emph{pStartY, pEndY $\in$ [-124, 120]} and \emph{pStartY $<$ pEndY}.
    \item $c_4$ is \emph{Word\_Dimension.Lemma = pTerms} where \emph{pTerms $\in$ \{t $\mid$ t $\in$ vocabulary \} }.
\end{itemize}

	Then, OLAP queries that extract top-$k$ keywords and documents bear the following contraints: $c_1 (Q1)$, $c_1 \wedge c_2 (Q2)$, $c_1 \wedge c_3 (Q3)$ and $c_1 \wedge c_2 \wedge c_3 (Q4)$; and $c_1 \wedge c_4 (Q1)$, $c_1 \wedge c_2 \wedge c_4 (Q2)$, $c_1 \wedge c_3 \wedge c_4 (Q3)$ and $c_1 \wedge c_2 \wedge c_3 \wedge c_4 (Q4)$, respectively.

\subsection{Performance Metrics and Execution Protocol}
\label{sec:MetricsProtocol}

	We use query response time as the only metric in both T$^2$K$^2$ and T$^2$K$^2$D$^2$. It is detail for each query as $t(Q_i ) \forall i \in [1, 4]$.  All queries $Q1$ to $Q4$ are executed 40 times for top-$k$ keywords and 10 times for top-$k$ documents, which is sufficient according to the central limit theorem. Average response times and standard deviations are computed for $t(Q_i)$. All executions are warm runs, i.e., either caching mechanisms must be deactivated, or a cold run of $Q1$ to $Q4$ must be executed once (but not taken into account in the benchmark's results) to fill in the cache. Queries must be written in the native scripting language of the target database system and executed directly inside said system using the command line interpreter.

\section{Benchmark Implementations}
\label{sec:Implementations}

\subsection{Weighting Schemes}
\label{sec:weightingSchemes}

	Given a corpus of documents $D = \{ d_1, d_2, ..., d_N\}$, where $N=|D|$ is the total number of documents in the dataset and $n$ the number of documents where some term $t$ appears. The TF-IDF weight is computed by multiplying the augmented term frequency $TF(t,d) = K + (1 - K) \cdot \frac{f_{t,d}}{\max_{t' \in d}(f_{t',d})}$) by the inverted document frequency $IDF(t,D) = 1 + \log\frac{N}{n}$, i.e., $TFIDF(t,d,D) = TF(t,d) \cdot IDF(t,D)$. The augmented form of $TF$ prevents a bias towards long tweets when the free parameter $K$ is set to $0.5$~\cite{Paltoglou2010}. It uses the number of co-occurrences $f_{t,d}$ of a word in a document, normalized with the frequency of the most frequent term $t'$, i.e., $\max_{t' \in d}(f_{t',d})$.
  
 	 The Okapi BM25 weight is given in Equation~\eqref{eq:okapi}, where $||d||$ is $d$'s length, i.e., the number of terms appearing in $d$. Average document length $avg_{d' \in D}(||d'||)$ is used to remove any bias towards long documents. The values of free parameters $k_1$ and $b$ are usually chosen, in absence of advanced optimization, as $k_1 \in [1.2,2.0]$ and $b=0.75$  \cite{Manning2008,SparckJones2000a,SparckJones2000b}.

 	\begin{equation}\label{eq:okapi}
      Okapi(t,d,D) = \frac{TFIDF(t,d,D) \cdot (k_1 + 1)}{TF(t,d) + k_1 \cdot (1 - b + b \cdot \frac{||d||}{ avg_{d' \in D}(||d'||)})}
  	\end{equation}

	To extract top-$k$ keywords, the overall relevance of a term $t$ for a given corpus $D$ is computed as the sum of all the TF-IDF (Equation~\eqref{eq:t_TFIDF}) or Okapi BM25 (Equation~\eqref{eq:t_okapi}) weights for that term.
    
\begin{equation}\label{eq:t_TFIDF}
	S_{TK}\_TFIDF(t,D) = \sum_{d_i \in D} TFIDF(t,d_i,D)
\end{equation}

\begin{equation}\label{eq:t_okapi} 
	S_{TK}\_Okapi(t,D) = \sum_{d_i \in D} Okapi(t,d_i,D)
\end{equation}
	   
    TF-IDF and Okapi BM25 are also used to ranks a set of documents based on the search query's terms appearing in each document. Given a search query $Q = \{ q_1, q_2, ..., q_m \}$, where $m=|Q|$ is the number of terms contained in the query, a document $d$ is scored by either summing all the TFDIF (Equation~\eqref{eq:r_TFIDF}) or the Okapi BM25 (Equation~\eqref{eq:r_okapi}) scores for the query terms in the document. 
    
	\begin{equation}\label{eq:r_TFIDF}
		S_{TD}\_TFIDF(Q,d,D) = \sum_{q_i \in Q} TF(q_i,d) \cdot IDF(q_i,D)
	\end{equation}

	\begin{equation}\label{eq:r_okapi}
		S_{TD}\_Okapi(Q,d,D) = \sum_{q_i \in Q} \frac{TFIDF(q_i,d,D) \cdot (k_1 + 1)}{TF(q_i,d) + k_1 \cdot (1 - b + b \cdot \frac{||d||}{ avg_{d' \in D}(||d'||)})}
	\end{equation}
	
\subsection{T$^2$K$^2$ implementation}
\label{sec:T2K2imp}

\subsubsection{Relational implementation}

	\paragraph{Dataset} 
    
\begin{figure}[!htbp]
    	\centering
        \includegraphics[width=\columnwidth]{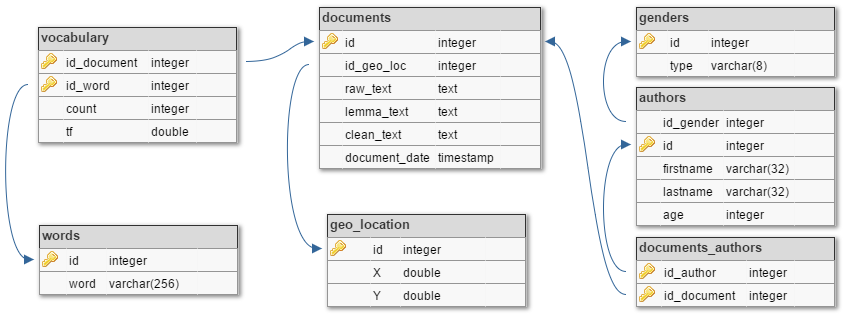}
        \caption{T$^2$K$^2$ relational data model} 
        \label{fig:T2K2dbschema}
    \end{figure}

    The logical relational schema for T$^2$K$^2$ used in both Oracle and PostgreSQL (Figure~\ref{fig:T2K2dbschema}) directly translates the conceptual schema from Figure~\ref{fig:T2K2ERDiagram}. 

	\paragraph{T$^2$K$^2$ queries}    
    
    The specification of T$^2$K$^2$'s top-$k$ keywords queries are expressed in relational algebra below.
    \begin{itemize}
        \item \emph{Q1 = $\gamma_L$( $\pi_{documents.id, words.word, f_w}$( $\sigma_{c_1}$( documents $\bowtie_{c_5}$ documents\_authors $\bowtie_{c_6}$ authors $\bowtie_{c_7}$ genders $\bowtie_{c_8}$ vocabulary $\bowtie_{c_9}$ words)))}.
        \item \emph{Q2 = $\gamma_L$( $\pi_{documents.id, words.word, f_w}$( $\sigma_{c_1 \wedge c_2}$( documents $\bowtie_{c_5}$ documents\_authors $\bowtie_{c_6}$ authors $\bowtie_{c_7}$ genders $\bowtie_{c_8}$ vocabulary $\bowtie_{c_8}$ words)))}.
        \item \emph{Q3 = $\gamma_L$( $\pi_{documents.id, words.word, f_w}$( $\sigma_{c_1 \wedge c_3}$( documents $\bowtie_{c_5}$ documents\_authors $\bowtie_{c_6}$ authors $\bowtie_{c_7}$ genders $\bowtie_{c_8}$ vocabulary $\bowtie_{c_9}$ words $\bowtie_{c_{10}}$ geo\_location)))}
        \item \emph{Q4 = $\gamma_L$( $\pi_{documents.id, words.word, f_w}$( $\sigma_{c_1 \wedge c_2 \wedge c_3}$( documents $\bowtie_{c_5}$ documents\_authors $\bowtie_{c_6}$ authors $\bowtie_{c_7}$ genders $\bowtie_{c_8}$ vocabulary $\bowtie_{c_9}$ words $\bowtie_{c_{10}}$ geo\_location)))}
    \end{itemize}
    
    The constraints for the queries are:
    \begin{itemize}
		\item $c_1$ is the constraint on gender.
    	\item $c_2$ is the constraint on time.
	    \item $c_3$ is the constraint on location.
        \item $c_5$ to $c_{10}$ are the join conditions as follows:
        \begin{itemize}
            \item $c_5$ is the JOIN condition between the \emph{documents} and \emph{documents\_authors} entities.
            \item $c_6$ is the JOIN condition between the  \emph{documents\_authors} and \emph{authors} entities.
            \item $c_7$ is the JOIN condition between the  \emph{authors} and \emph{genders} entities.
            \item $c_8$ is the JOIN condition between the \emph{documents} and \emph{vocabulary} entities.
            \item $c_9$ is the JOIN condition between the  \emph{vocabulary} and \emph{words} entities.
            \item $c_{10}$ is the JOIN condition between the \emph{documents} and \emph{geo\_location}.
        \end{itemize}
     \end{itemize}
	     
	Function $f_w$ is used to compute the weighting schema using nested queries:
	\begin{itemize}
		\item TF-IDF with the parameters $vocabulary.tf = TF(t,d)$, the total number of tweets in the corpus and the number of tweets where a term appears; the last two parameters are computed using individual nested queries.
		\item Okapi BM25 with parameters $vocabulary.tf = TF(t,d)$, the total number of tweets in the corpus, the length of each tweet and the number of tweets where a term appears; the last three parameters are computed using individual nested queries.
	\end{itemize}

	The last operator is the aggregation operator $\gamma_L$ where $L=(F, G)$ is:
    \begin{itemize}
		\item \emph{F = sum($f_w$)}, the \emph{sum} is the aggregation function that computes $S_{TK}\_TFIDF(t,D)$ (Equation~\eqref{eq:t_TFIDF}), respectively $S_{TK}\_Okapi(t,D)$ ((Equation~\eqref{eq:t_okapi})).
        \item $G = (words.word)$ is a list of attributes in the GROUP BY clause, in the case the terms (\emph{words.word}).
	\end{itemize}
	
     The queries for determining the top-$k$ documents are similar to the ones that compute the top-$k$ keywords, i.e., the join conditions and function  $f_w$ are the same. Only the constraint $c_4$ is added to select the tweets that contain the required search terms. The modification appears at the aggregation operator $\gamma_L$ where the list of the GROUP BY clause attributes changes, this time grouping is done by using the $documents.id$. Moreover, the aggregation \emph{F = sum($f_w$)} computes the hierarchy of tweets, $S_{TD}\_TFIDF(Q,d,D)$  (Equation~\eqref{eq:r_TFIDF}) for TF-IDF or $S_{TD}\_Okapi(Q,d,D)$ (Equation~\eqref{eq:r_okapi}) for Okapi BM25, where $Q$ is the list of search terms. T$^2$K$^2$'s top-$k$ documents queries expressed in relational algebra are:
    \begin{itemize}
        \item \emph{Q1 = $\gamma_L$( $\pi_{documents.id, words.word, f_w}$( $\sigma_{c_1 \wedge c_4}$( documents $\bowtie_{c_5}$ documents\_authors $\bowtie_{c_6}$ authors $\bowtie_{c_7}$ genders $\bowtie_{c_8}$ vocabulary $\bowtie_{c_9}$ words)))}.
        \item \emph{Q2 = $\gamma_L$( $\pi_{documents.id, words.word, f_w}$( $\sigma_{c_1 \wedge c_2 \wedge c_4}$( documents $\bowtie_{c_5}$ documents\_authors $\bowtie_{c_6}$ authors $\bowtie_{c_7}$ genders $\bowtie_{c_8}$ vocabulary $\bowtie_{c_8}$ words)))}.
        \item \emph{Q3 = $\gamma_L$( $\pi_{documents.id, words.word, f_w}$( $\sigma_{c_1 \wedge c_3 \wedge c_4}$( documents $\bowtie_{c_5}$ documents\_authors $\bowtie_{c_6}$ authors $\bowtie_{c_7}$ genders $\bowtie_{c_8}$ vocabulary $\bowtie_{c_9}$ words $\bowtie_{c_{10}}$ geo\_location)))}
        \item \emph{Q4 = $\gamma_L$( $\pi_{documents.id, words.word, f_w}$( $\sigma_{c_1 \wedge c_2 \wedge c_3 \wedge c_4}$( documents $\bowtie_{c_5}$ documents\_authors $\bowtie_{c_6}$ authors $\bowtie_{c_7}$ genders $\bowtie_{c_8}$ vocabulary $\bowtie_{c_9}$ words $\bowtie_{c_{10}}$ geo\_location)))}
    \end{itemize}

\subsubsection{Document-ortiented implementation}
    
    \paragraph{Dataset} 
    
    In a Document Oriented Database Management System (DODBMS), all information is typically stored in a single collection. An example of DODBMS document is presented in Figure~\ref{fig:docexample}. The many-to-many \emph{Vocabulary} relationship from Figure~\ref{fig:T2K2ERDiagram} is modeled as a nested document for each record. The information about user and date become single fields in a document, while the location becomes an array. 
   
\begin{figure}[!hbtp]
\centering      
\begin{tabular}{|c|}
\hline
\begin{lstlisting}[language=Java]
{   _id : 644626677310603264, 
    rawText : "Amanda's car is too much for my headache", 
    cleanText : "Amanda is car is too much for my headache", 
    lemmaText : "amanda car headache", 
    author : 970993142, 
    geoLocation : [ 32, 79 ], 
    gender : "male", 
    age : 23,
    lemmaTextLength : 3,
    words : [ { "tf" : 1, "count" : 1, "word" : "amanda"}, 
            { "tf" : 1, "count" : 1, "word" : "car" }, 
            { "tf" : 1, "count" : 1, "word" : "headache"} ], 
    date : ISODate("2015-09-17T23:39:11Z") }
\end{lstlisting} \\ \hline
\end{tabular}
\caption{Sample T$^2$K$^2$ DODBMS Document}
\label{fig:docexample}
\end{figure}
    
    \paragraph{Queries} 
    
    In DODBMSs, user-defined functions written in JavaScript are used to compute top-$k$ keywords, regardless of the schema. For T$^2$K$^2$, the TF-IDF weight can take advantage of both native database aggregation (aggregation pipeline - AP) and MapReduce (MR). However, due to the multitude of parameters involvedand the calculations needed for the Okapi BM25 weighting scheme, the NA method is usually difficult to develop. Thus, we recommend to only use MR in benchmark runs.
    
\subsection{T$^2$K$^2$D$^2$ implementation}

\subsubsection{Relational implementation}

    \paragraph{Dataset} 
    
\begin{figure}[!htbp]
    	\centering
        \includegraphics[width=\columnwidth]{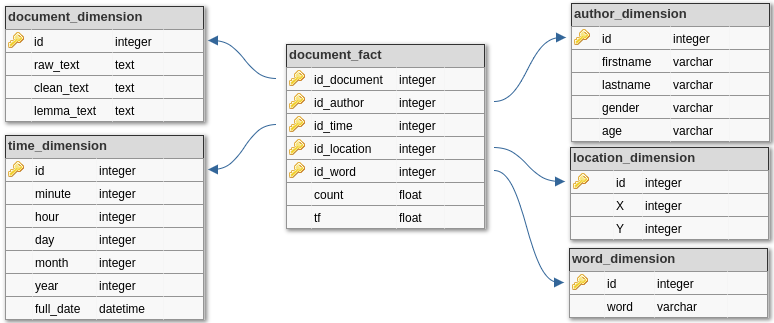}
        \caption{Star schema diagram} 
        \label{fig:starschema}
	\end{figure}

    The logical multidimensional star schema for T$^2$K$^2$D$^2$ used in both relational databases management systems (Figure~\ref{fig:starschema}) directly translates from the conceptual schema from Figure~\ref{fig:ERStarDiagram}. 

    \paragraph{Queries} 
    
    T$^2$K$^2$D$^2$'s top-$k$ keywords queries expressed in relational algebra are:
    \begin{itemize}
        \item \emph{Q1 = $\gamma_L$( $\pi_{word\_dimension.word, f_w}$( $\sigma_{c_1}$( document\_fact $\bowtie_{c_5}$ word\_dimension $\bowtie_{c_6}$ author\_dimension )))}.
        \item \emph{Q2 = $\gamma_L$( $\pi_{word\_dimension.word, f_w}$( $\sigma_{c_1 \wedge c_2}$( document\_fact $\bowtie_{c_5}$ word\_dimension $\bowtie_{c_6}$ author\_dimension $\bowtie_{c_7}$ time\_dimension )))}.
        \item \emph{Q3 = $\gamma_L$( $\pi_{word\_dimension.word, f_w}$( $\sigma_{c_1 \wedge c_3}$( document\_fact $\bowtie_{c_5}$ word\_dimension $\bowtie_{c_6}$ author\_dimension $\bowtie_{c_8}$ location\_dimension )))}.
        \item \emph{Q4 = $\gamma_L$( $\pi_{word\_dimension.word, f_w}$( $\sigma_{c_1 \wedge c_2 \wedge c_3}$( document\_fact $\bowtie_{c_5}$ word\_dimension $\bowtie_{c_6}$ author\_dimension $\bowtie_{c_7}$ time\_dimension $\bowtie_{c_8}$ location\_dimension )))}.
    \end{itemize}
   
	The constraints for the queries are:
    \begin{itemize}
    	\item $c_1$ is the constraint on gender.
    	\item $c_2$ is the constraint on time.
	    \item $c_3$ is the constraint on location.
        \item $c_5$ to $c_8$ are the JOIN constraints:
        \begin{itemize}
            \item $c_5$ is the JOIN constraint between \emph{document\_fact} and \emph{word\_dimension} entities.
            \item $c_6$ is the JOIN constraint between \emph{document\_fact} and \emph{author\_dimension} entities.
            \item $c_7$  is the JOIN constraint between \emph{document\_fact} and \emph{time\_dimension} entities.
            \item $c_8$  is the JOIN constraint between \emph{document\_fact} and \emph{location\_dimension} entities.            
        \end{itemize}
	\end{itemize}
	
    Function $f_w$ is used to compute the weighting schema using nested queries:
    \begin{itemize}
		\item TF-IDF with the parameters $document\_fact.tf = TF(t,d)$, the total number of tweets in the corpus and the number of tweets where a term appears; the last two parameters are computed using individual nested queries.
		\item Okapi BM25 with parameters $document\_fact.tf = TF(t,d)$, the total number of tweets oin the corpus, the length of each tweet and the number of tweets where a term appears; the last three parameters are computed using individual nested queries.
	\end{itemize}
	
	Finally, we present the aggregation operator $\gamma_L$, where $L=(F, G)$ is:
    \begin{itemize}
            \item \emph{F = sum($f_w$)}, the \emph{sum} is the aggregation function that computes $S_{TK}\_TFIDF(t,D)$ (Equation~\eqref{eq:t_TFIDF}), respectively $S_{TK}\_Okapi(t,D)$ ((Equation~\eqref{eq:t_okapi})).
            \item $G = (word\_dimension.word)$ is a list of attributes in the GROUP BY clause, in the case the terms (\emph{word\_dimension.word}).
    \end{itemize}

    T$^2$K$^2$D$^2$'s top-$k$ keywords queries expressed in relational algebra are:
	\begin{itemize}
		\item \emph{Q1 = $\gamma_L$( $\pi_{document\_fact.id\_document, f_w}$( $\sigma_{c_1 \wedge c_4}$( document\_fact $\bowtie_{c_5}$ word\_dimension $\bowtie_{c_6}$ author\_dimension )))}.
		\item \emph{Q2 = $\gamma_L$( $\pi_{document\_fact.id\_document, f_w}$( $\sigma_{c_1 \wedge c_2 \wedge c_4}$( document\_fact $\bowtie_{c_5}$ word\_dimension $\bowtie_{c_6}$ author\_dimension $\bowtie_{c_7}$ time\_dimension )))}.
		\item \emph{Q3 = $\gamma_L$( $\pi_{document\_fact.id\_document, f_w}$( $\sigma_{c_1 \wedge c_3 \wedge c_4}$( document\_fact $\bowtie_{c_5}$ word\_dimension $\bowtie_{c_6}$ author\_dimension $\bowtie_{c_8}$ location\_dimension )))}.
		\item \emph{Q4 = $\gamma_L$( $\pi_{document\_fact.id\_document, f_w}$( $\sigma_{c_1 \wedge c_2 \wedge c_3 \wedge c_4}$( document\_fact $\bowtie_{c_5}$ word\_dimension $\bowtie_{c_6}$ author\_dimension $\bowtie_{c_7}$ time\_dimension $\bowtie_{c_8}$ location\_dimension )))}.
	\end{itemize}
    
    T$^2$K$^2$D$^2$ queries that compute the top-$k$ documents on the star schema are similar with the ones that compute the top-$k$ keywords, only the constraint $c_4$ is added to select the documents that contain the required search terms. Regardless of the weighting schema used, the $f_w$ function is computed differently because it uses different parameters, as follows:
	\begin{itemize}
		\item $document\_fact.tf = TF(t,d)$.
        \item the length of each tweet is computed using a nested query
        \item the number of tweets where a term appears is computed using a nested query
	\end{itemize}
    
    Moreover, the aggregation operator $\gamma_L$ is also different from the one in the top-$k$ Keywords, i.e., the grouping is done using the tweet unique identifier ($document\_fact.id\_document$) and the aggregation function \emph{F = sum($f_w$)} computes the hierarchy of tweets using $S_{TD}\_TFIDF(Q,d,D)$ (Equation~\eqref{eq:r_TFIDF}) for TF-IDF, respectively  $S_{TD}\_Okapi(Q,d,D)$ (Equation~\eqref{eq:r_okapi}) for Okapi BM25, where $Q$ is the list of search terms.
    
\subsubsection{Document-ortiented implementation}

\paragraph{Dataset}  

	To implement T$^2$K$^2$D$^2$'s star schema in a DODBMS each dimension becomes a nested document, thus a document contains multiple nested documents~\cite{Chevalier2015}, one for each dimension. Figure~\ref{fig:StarDocEx} presents a document sample for implementing a multidimensional model in a DODBMS.

\begin{figure}[!htbp]
\centering      
\begin{tabular}{|c|}
\hline
\begin{lstlisting}[language=Java]
{   "_id" : ObjectId("595c8d705f32f5b46802b24c"),
    "document_dimension" : {
      "id_document" : 644626677310603264,
      "rawText" : "Amanda's car is too much for my headache",
      "cleanText" : "Amanda is car is too much for my headache",
      "lemmaText" : "amanda car headache",
      "lemmaTextLength" : 3 },
    "author_dimension" : {
      "author" : 970993142,
      "age" : 23,
      "gender" : "male" },
    "time_dimension" : {
      "minute" : 39, "hour" : 23, 
      "day" : 17, "month" : 9, "year" : 2015,
      "full_date" : ISODate("2015-09-17T23:39:11Z") },
    "location_dimension" : { "X" : 32, "Y" : 79 },
    "words" : [ { "tf" : 1, "count" : 1, "word" : "amanda" },
                { "tf" : 1, "count" : 1, "word" : "car" },
                { "tf" : 1, "count" : 1, "word" : "headache" } ] }
\end{lstlisting} \\ \hline
\end{tabular}
\caption{Sample T$^2$K$^2$D$^2$ DODBMS Document}
\label{fig:StarDocEx}
\end{figure}    

\paragraph{Queries}   
   
   To compute the top-$k$ keywords using T$^2$K$^2$D$^2$ implementation and to compute the top-$k$ documents, regardless of the schema, only the MR implementation is used. Also, the benchmark tests are implemented in the DODBMS native JavaScript language.

\section{Experiments}
\label{sec:Experiments}

\subsection{Experimental Conditions}
\label{sec:ExpConditions}

    All tests run on an IBM System x3550 M4 with 64GB of RAM and an Intel(R) Xeon(R) CPU E5-2670 v2 @ 2.50GHz. The database systems we use are MongoDB v3.4.6, Oracle v12.1.0.2.0 and PostgreSQL v9.6.1. Initially, we planned to benchmark CouchDB 2.0.0, too. Unfortunately, we found no in-database solution to implement T$^2$K$^2$'s queries, because of CouchDB's limited programming environment and MR framework. The code of all Oracle and PostgreSQL queries and MongoDB functions, together with benchmarking results, are available on Github\footnote{\url{https://github.com/cipriantruica/T2K2D2_Benchmark}}.
            
 	MongoDB's performance is tested both in a single-instance and a distributed environment. The distributed environment is composed of a configuration node and 5 slave nodes; each slave becoming a worker in the MR algorithm. The database is distributed using horizontal scaling to test whether weighting scheme computing time decreases as the number of nodes increases. In MongoDB, horizontal scaling is achieved using sharding, and each individual data partition is refered as a shard~\cite{Bonnet2011}.
 	    
	The relational database implementations are tested only on a single-instance environment, because the Oracle Real Application Cluster that is needed to distribute Oracle databases is not free; and because PostgreSQL does not distribute natively, requiring third party software. 

	Besides indexes on primary keys, no other indexes are used in our tests. The sharding key in the MongoDB distributed environment is the tweet's unique identifier. 

	Eventually, query parameterization is provided in Table~\ref{tbl:qparam}.

    \begin{table}[!ht]
        \centering
        \small
        \caption{Query parameter values}
        \label{tbl:qparam}
        \begin{tabular}{|c|c|c|c|c|c|c|c|}
            \hline
            \textbf{\textit{pGender}} & \textbf{\textit{pStartDate}} & \textbf{\textit{pEndDate}} & \textbf{\textit{pStartX}} & \textbf{\textit{pEndX}} & \textbf{\textit{pStartY}} & \textbf{\textit{pEndY}}  & \textbf{\textit{pWords}} \\ \hline
            \begin{tabular}[c]{@{}c@{}}male $|$\\ female\end{tabular} & \begin{tabular}[c]{@{}c@{}}2015-09-17\\ 00:00:00\end{tabular} & \begin{tabular}[c]{@{}c@{}}2015-09-18\\ 00:00:00\end{tabular} & 20 & 40 & -100 & 100 & \multicolumn{1}{c|}{\begin{tabular}[c]{@{}c@{}}think $|$\\ today $|$\\ friday\end{tabular}} \\ \hline
        \end{tabular}
    \end{table}
    
\subsection{Dataset}

 	Experiments are done on a 2\,500\,000 tweets corpus. The initial corpus is split into 5 different datasets that all keep an equal balance between the number of tweets for both genders, location and date. These datasets contain 500\,000, 1\,000\,000, 1\,500\,000, 2\,000\,000 and 2\,500\,000 tweets, respectively. They allow scaling experiments and are associated to a scale factor ($SF$) parameter, where $SF \in \{0.5, 1, 1.5, 2, 2.5\}$, for conciseness sake. 

\subsubsection{Query Selectivity}

	Selectivity, i.e., the amount of retrieved data ($n(Q)$)  w.r.t. the total amount of data available ($N$), depends on the number of attributes in the where and group by clauses. The selectivity formula used for a query $Q$ is $S(Q) = 1-\frac{n(Q)}{N}$. 
    
    For T$^2$K$^2$, all queries traverse the \emph{Document}, \emph{Vocabulary}, \emph{Word} and \emph{Gender} entities, and the \emph{Write} and \emph{Vocabulary} relationships, to extract lemmas and compute weights. All queries use the the \emph{document\_fact}, \emph{Word\_Dimnesion} and \emph{Author\_Dimension} for T$^2$K$^2$D$^2$.
    
    All queries filter on gender, to determine the trending words for female and male users. Starting from $Q1$, subsequent queries $Q2$ to $Q4$ are built by decreasing selectivity (Table~\ref{tbl:selectivity}), regardless of the benchmark, i.e., T$^2$K$^2$ or T$^2$K$^2$D$^2$. Moreover, by adding a constraint on the location in $Q3$ and $Q4$, query complexity changes, too.

\begin{table}[!ht]
        \centering
        \caption{Top-$k$ keywords query selectivity}
        \label{tbl:selectivity}
        \begin{tabular}{|c|c|c|c|c|c|c|c|c|}
            \hline
            \multicolumn{1}{|c|}{\textbf{\textit{SF}}} & \multicolumn{1}{c|}{\textbf{\begin{tabular}[c]{@{}l@{}}\textit{Q1}\\ male\end{tabular}}} & \multicolumn{1}{c|}{\textbf{\begin{tabular}[c]{@{}l@{}}\textit{Q1}\\ female\end{tabular}}} & \multicolumn{1}{c|}{\textbf{\begin{tabular}[c]{@{}l@{}}\textit{Q2}\\ male\end{tabular}}} & \multicolumn{1}{c|}{\textbf{\begin{tabular}[c]{@{}l@{}}\textit{Q2}\\ female\end{tabular}}} & \multicolumn{1}{c|}{\textbf{\begin{tabular}[c]{@{}l@{}}\textit{Q3}\\ male\end{tabular}}} & \multicolumn{1}{c|}{\textbf{\begin{tabular}[c]{@{}l@{}}\textit{Q3}\\ female\end{tabular}}} & \multicolumn{1}{c|}{\textbf{\begin{tabular}[c]{@{}l@{}}\textit{Q4}\\ male\end{tabular}}} & \multicolumn{1}{c|}{\textbf{\begin{tabular}[c]{@{}l@{}}\textit{Q4}\\ female\end{tabular}}} \\ \hline
            0.5 & 0.336 & 0.337 & 0.517 & 0.517 & 0.556 & 0.558 & 0.677 & 0.679  \\ \hline
            1   & 0.342 & 0.342 & 0.662 & 0.662 & 0.562 & 0.565 & 0.774 & 0.775  \\ \hline
            1.5 & 0.347 & 0.346 & 0.736 & 0.736 & 0.569 & 0.572 & 0.823 & 0.824  \\ \hline
            2   & 0.351 & 0.350 & 0.783 & 0.783 & 0.574 & 0.575 & 0.855 & 0.856  \\ \hline
            2.5 & 0.353 & 0.354 & 0.815 & 0.815 & 0.579 & 0.580 & 0.876 & 0.877  \\ \hline
        \end{tabular}
    \end{table}

	Selectivity for the top-$k$ documents is decreased even more by adding a condition on the words attribute for all the queries.  Moreover, for both benchmarks, selectivity remains the same (Table~\ref{tbl:selectivity_docs_3w}). 

    \begin{table}[!ht]
        \centering
        \caption{Top-$k$ documents selectivity for 3 search terms}
        \label{tbl:selectivity_docs_3w}
        \begin{tabular}{|c|c|c|c|c|c|c|c|c|}
            \hline
            \textbf{\textit{SF}} & \multicolumn{1}{l|}{\textbf{\begin{tabular}[c]{@{}l@{}}\textit{Q1}\\ male\end{tabular}}} & \multicolumn{1}{l|}{\textbf{\begin{tabular}[c]{@{}l@{}}\textit{Q1}\\ female\end{tabular}}} & \multicolumn{1}{l|}{\textbf{\begin{tabular}[c]{@{}l@{}}\textit{Q2}\\ male\end{tabular}}} & \multicolumn{1}{l|}{\textbf{\begin{tabular}[c]{@{}l@{}}\textit{Q2}\\ female\end{tabular}}} & \multicolumn{1}{l|}{\textbf{\begin{tabular}[c]{@{}l@{}}\textit{Q3}\\ male\end{tabular}}} & \multicolumn{1}{l|}{\textbf{\begin{tabular}[c]{@{}l@{}}\textit{Q3}\\ female\end{tabular}}} & \multicolumn{1}{l|}{\textbf{\begin{tabular}[c]{@{}l@{}}\textit{Q4}\\ male\end{tabular}}} & \multicolumn{1}{l|}{\textbf{\begin{tabular}[c]{@{}l@{}}\textit{Q5}\\ female\end{tabular}}} \\ \hline
        0.5 & 0.9844 & 0.9848 & 0.9904 & 0.9905 & 0.9921 & 0.9926 & 0.9951 & 0.9954 \\ \hline
        1   & 0.9866 & 0.9868 & 0.9952 & 0.9953 & 0.9932 & 0.9936 & 0.9975 & 0.9977 \\ \hline
        1.5 & 0.9835 & 0.9837 & 0.9968 & 0.9968 & 0.9917 & 0.9920 & 0.9984 & 0.9985 \\ \hline
        2   & 0.9822 & 0.9824 & 0.9976 & 0.9976 & 0.9910 & 0.9913 & 0.9988 & 0.9988 \\ \hline
        2.5 & 0.9825 & 0.9827 & 0.9981 & 0.9981 & 0.9912 & 0.9915 & 0.9990 & 0.9991 \\ \hline
    \end{tabular}
    \end{table}
    
\subsubsection{Query Complexity}

	Complexity relates to the number of traversals involved in the query. Query complexity depends on the number of relationship and entity traversals. 
    
    Independently from any weighting schema, for T$^2$K$^2$ computing the number of tweets where a term appears and the total number of documents implies two aggregation queries that traverse either all entities ($Q3$, $Q4$) or all entities but \emph{Geo\_Location} ($Q1$, $Q2$), resulting in a number of 5 traversals for $Q3$ and $Q4$ and 4 traversals for $Q1$ and $Q2$. Moreover, some weighting schemes may require additional queries, e.g., Okapi BM25 involves tweet lengths. Table~\ref{tbl:query_complexity} presents query complexities w.r.t. weighting schemes.
   Query complexity is the same for T$^2$K$^2$ implementation regardless of computing the top-$k$ keywords or top-$k$ documents. 
   
       \begin{table}[!ht]
        \centering
        \caption{T$^2$K$^2$ query complexity}
        \label{tbl:query_complexity}
        \begin{tabular}{l|c|c|c|c|}
            \cline{2-5}
            & \textbf{\textit{Q1}} & \textbf{\textit{Q2}} & \textbf{\textit{Q3}} & \textbf{\textit{Q4}} \\ \hline
            \multicolumn{1}{|l|}{\textbf{TF-IDF}}      & 12          & 12          & 15          & 15          \\ \hline
            \multicolumn{1}{|l|}{\textbf{Okapi BM25}} & 17          & 17          & 21          & 21          \\ \hline
        \end{tabular}
    \end{table} 
    
   Table~\ref{tbl:query_complexity_star} present the T$^2$K$^2$D$^2$ queries complexity for top-$k$ keywords.
   
    \begin{table}[!ht]
        \centering
        \caption{T$^2$K$^2$D$^2$ query complexity for top-$k$ keywords}
        \label{tbl:query_complexity_star}
        \begin{tabular}{l|c|c|c|c|}
            \cline{2-5}
            & \textbf{\textit{Q1}} & \textbf{\textit{Q2}} & \textbf{\textit{Q3}} & \textbf{\textit{Q4}} \\ \hline
            \multicolumn{1}{|l|}{\textbf{TF-IDF}}      & 3          & 5          & 5          & 7          \\ \hline
            \multicolumn{1}{|l|}{\textbf{Okapi BM25}} & 4          & 6          & 6          & 8          \\ \hline
        \end{tabular}
    \end{table}
    
    The T$^2$K$^2$D$^2$ queries complexity for top-$k$ documents is presented in Table~\ref{tbl:query_complexity_star_docs}.
    
    \begin{table}[!ht]
        \centering
        \caption{T$^2$K$^2$D$^2$ query complexity for top-$k$ documents}
        \label{tbl:query_complexity_star_docs}
        \begin{tabular}{l|c|c|c|c|}
            \cline{2-5}
            & \textbf{\textit{Q1}} & \textbf{\textit{Q2}} & \textbf{\textit{Q3}} & \textbf{\textit{Q4}} \\ \hline
            \multicolumn{1}{|l|}{\textbf{TF-IDF}}      & 5        & 8        & 8       & 11          \\ \hline
            \multicolumn{1}{|l|}{\textbf{Okapi BM25}} & 6        & 9        & 9       & 12          \\ \hline
        \end{tabular}
    \end{table}

\subsection{Experimental Results}
\label{sec:ExpResults}

\subsubsection{T$^2$K$^2$}

	\paragraph{Weighting schema comparison}

\begin{figure}[!htbp]
    \centering
    \begin{subfigure}{\columnwidth}
      \centering
				\includegraphics[width=\columnwidth]{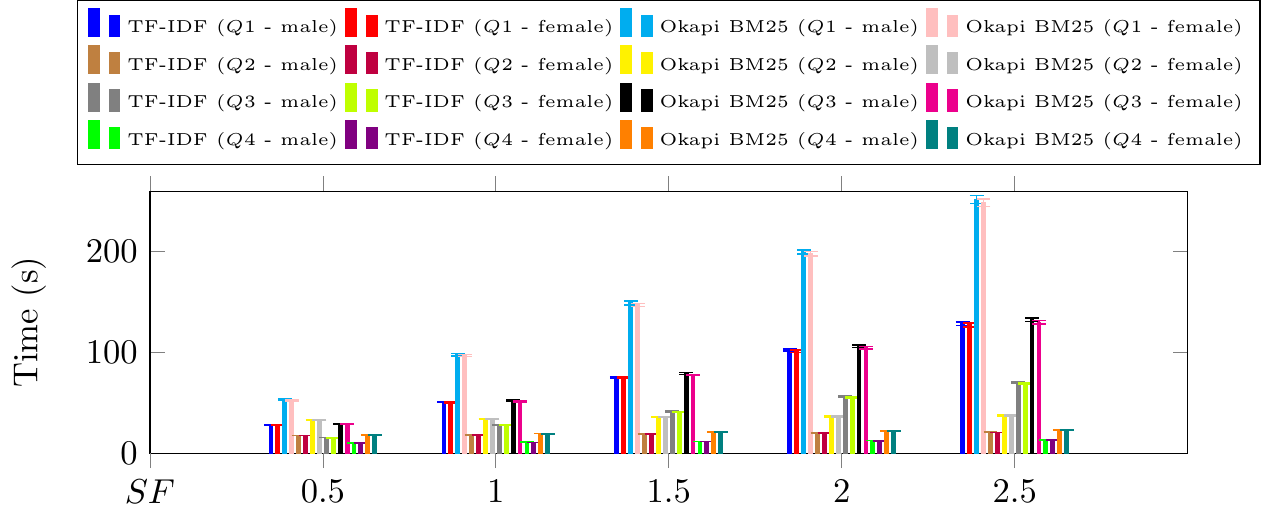}
            \caption{MongoDB MR: TF-IDF vs. Okapi BM25}
            \label{fig:TFIDF_okapi_MR}
    \end{subfigure}

    \begin{subfigure}{\columnwidth}
       \centering
        \includegraphics[width=\columnwidth]{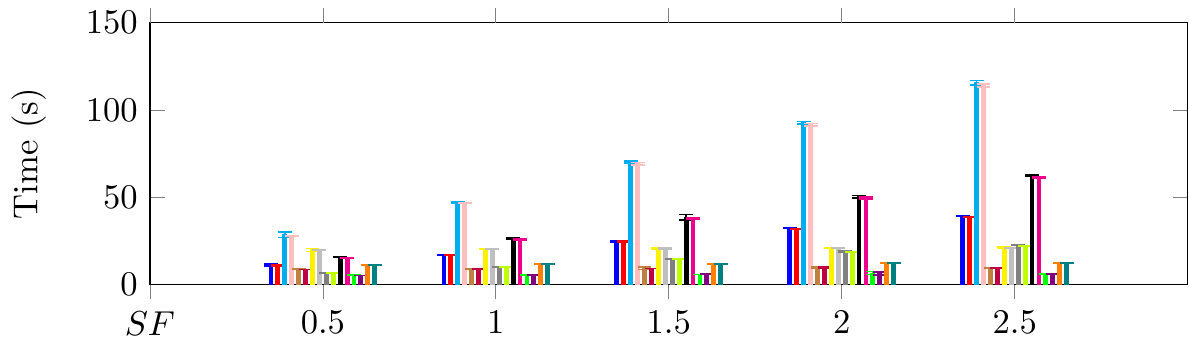}
        \caption{MongoDB MR Distributed: TF-IDF vs. Okapi BM25}
        \label{fig:TFIDF_okapi_MR_dist}
    \end{subfigure}    
    \begin{subfigure}{\columnwidth}
      \centering
        \includegraphics[width=\columnwidth]{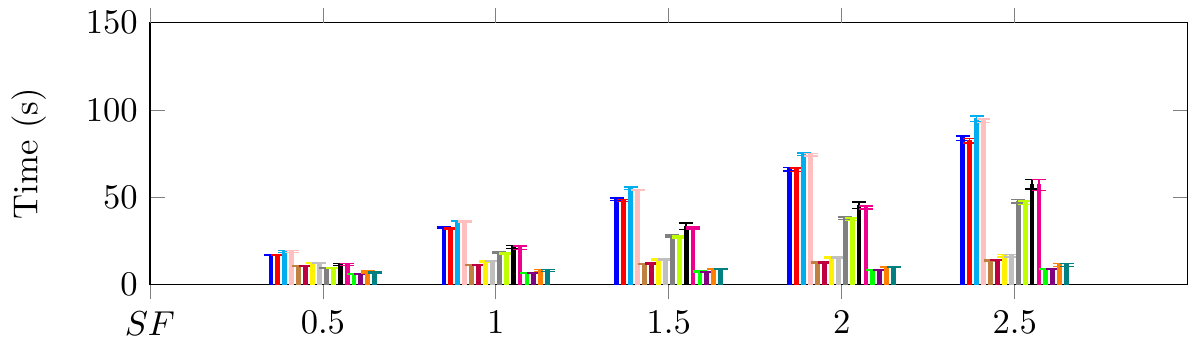}
        \caption{Oracle: TF-IDF vs. Okapi BM25}
        \label{fig:TFIDF_okapi_Oracle}
    \end{subfigure}   
    \begin{subfigure}{\columnwidth}
       \centering
				\includegraphics[width=\columnwidth]{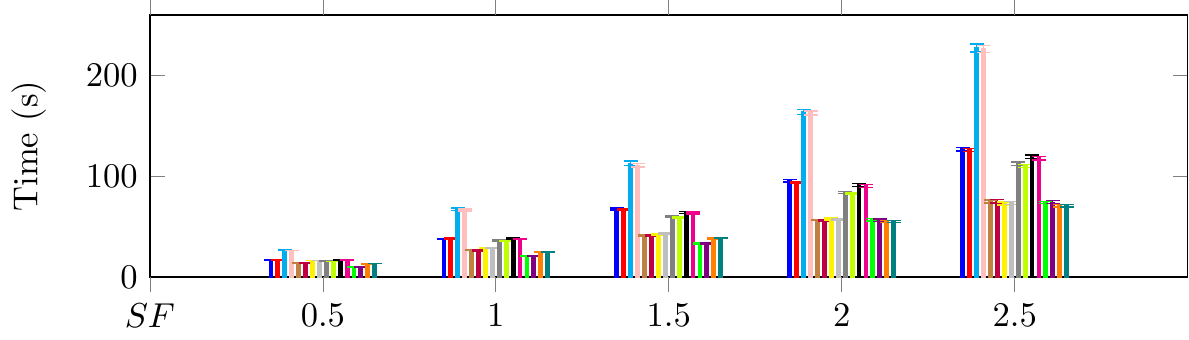}
            \caption{PostgreSQL: TF-IDF vs. Okapi BM25}
            \label{fig:TFIDF_okapi_PostgreSQL}
    \end{subfigure}
    \caption{Top-$k$ keywords TF-IDF vs. Okapi BM25 comparison}
    \label{fig:TFIDF_okapi}
\end{figure}

	Figure~\ref{fig:TFIDF_okapi} presents a comparison by database system and weighting scheme of query response time for retrieving top-$k$ keywords w.r.t. scale factor $SF$. This comparison uses only MongoDB's MR query implementations in  single-instance and distributed environments, because there is no AP implementation for queries that use Okapi BM25.

	In MongoDB, computing  TF-IDF is faster than computing Okapi BM25  for all queries, regardless of $SF$  (Figure~\ref{fig:TFIDF_okapi_MR}). The same results are observed in the distributed environment (Figure~\ref{fig:TFIDF_okapi_MR_dist}). The biggest difference in computation time between TF-IDF and Okapi BM25 is obtained for $Q1$, while the smallest is obtained for $Q4$ because of its  higher selectivity. 
        
    The execution times of $Q2$ and $Q4$ on Oracle and PostgreSQL are almost constant for all tested scenarios, regardless of weighting scheme or $SF$, because selectivity is constant in these queries. TF-IDF computing time is almost a third of the computing time for Okapi BM25's for $Q4$ in Oracle (Figure~\ref{fig:TFIDF_okapi_Oracle}) due to the complexity of the query itself and of the formula for computing Okapi BM25 w.r.t. TF-IDF.  In PostgreSQL, the biggest performance difference between TF-IDF and Okapi BM25 is obtained for $Q1$ (Figure~\ref{fig:TFIDF_okapi_PostgreSQL}) due to the large number of rows returned when joining all tables.   	

\begin{figure}[!htbp]
    \centering
    \begin{subfigure}{\columnwidth}
      \centering
				\includegraphics[width=\columnwidth]{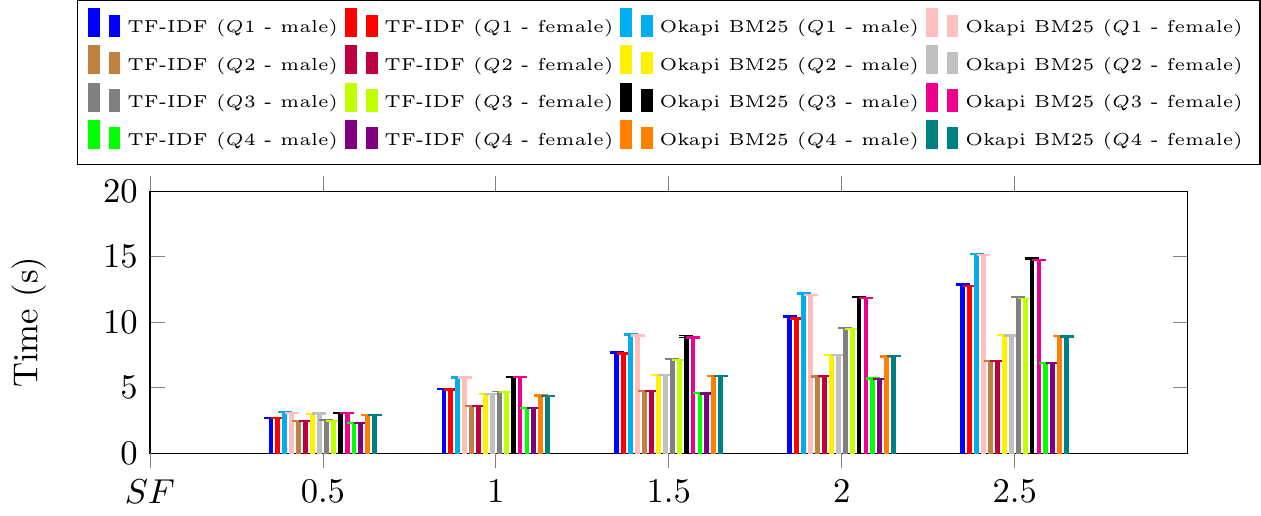}
            \caption{MongoDB MR: TF-IDF vs. Okapi BM25}
            \label{fig:TFIDF_okapi_MR_docs}
    \end{subfigure}
    \begin{subfigure}{\columnwidth}
       \centering
        \includegraphics[width=\columnwidth]{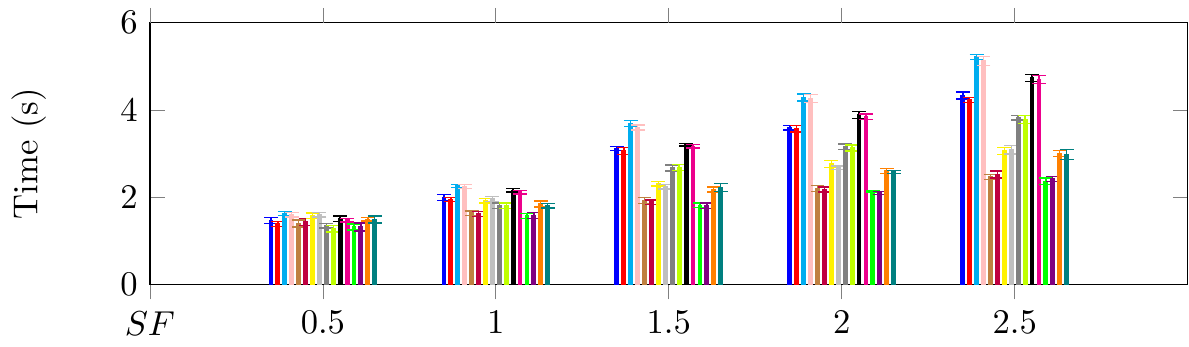}
        \caption{MongoDB MR Distributed: TF-IDF vs. Okapi BM25}
        \label{fig:TFIDF_okapi_MR_dist_docs}
    \end{subfigure}    
    \begin{subfigure}{\columnwidth}
      \centering
		  \includegraphics[width=\columnwidth]{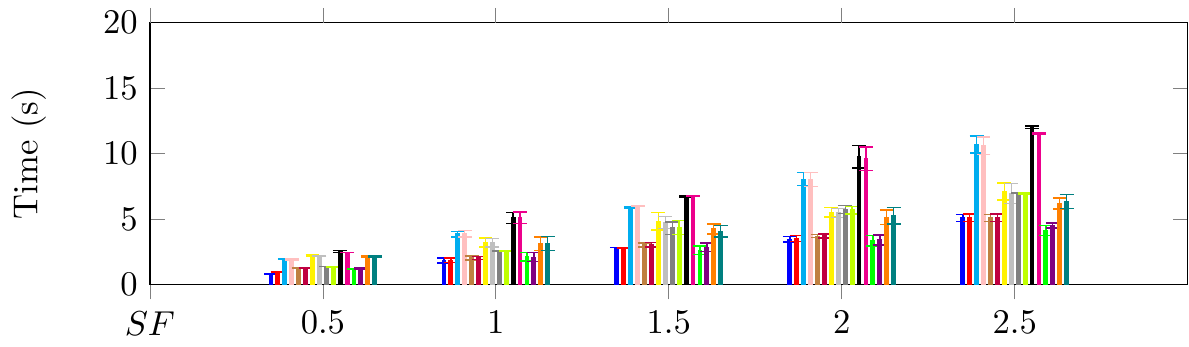}
        \caption{Oracle: TF-IDF vs. Okapi BM25}
        \label{fig:TFIDF_okapi_Oracle_docs}
    \end{subfigure}   
    \begin{subfigure}{\columnwidth}
       \centering
				\includegraphics[width=\columnwidth]{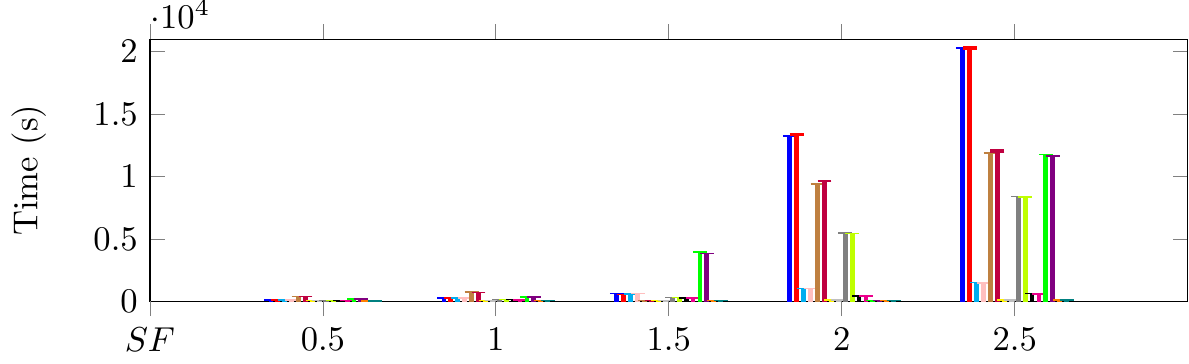}
            \caption{PostgreSQL: TF-IDF vs. Okapi BM25}
            \label{fig:TFIDF_okapi_PostgreSQL_docs}
    \end{subfigure}
    \caption{Top-$k$ documents TF-IDF vs. Okapi BM25 comparison}
    \label{fig:TFIDF_okapi_docs}
\end{figure}

	Figure~\ref{fig:TFIDF_okapi_docs} presents a comparison by database system and weighting scheme of query response time for retrieving top-$k$ documents w.r.t. scale factor $SF$. 

	In MongoDB, computing  TF-IDF is faster than computing Okapi BM25 for all queries, regardless of $SF$ (Figure~\ref{fig:TFIDF_okapi_MR_docs}). The same results are observed in the distributed environment (Figure~\ref{fig:TFIDF_okapi_MR_dist_docs}). In this case the performance of MongoDB is stable, regardless of the environment (single node or distributed), the performance curve increases linearly (Figures~\ref{fig:TFIDF_okapi_MR_docs} and~\ref{fig:TFIDF_okapi_MR_dist_docs}). Such results are an expected consequence of the variation of query selectivity and complexity, as seletivity decreases and complexity increases from $Q1$ to $Q2$. Furthermore, as $SF$ increases, so does seletivity in all queries.
        
	The execution times of $Q1$ and $Q3$ for Oracle are getting worse when $SF$ increases, since selectivity increases with the number of documents. In Oracle, the execution time for computing TF-IDF is half of the execution time for computing Okapi BM25 (Figure~\ref{fig:TFIDF_okapi_Oracle_docs}) regardless of $SF$. This is a direct consequence of query complexity. The complexity of queries that use the TF-IDF weighting is indeed lower than those that use Okapi BM25 (Table~\ref{tbl:query_complexity}). Moreover, in Oracle, the execution time is unstable as the standard deviation fluctuates w.r.t $SF$ due to automatic cache cleaning when query results are too big to be stored in the database's internal buffers.
	
	In PostgreSQL, the biggest performance difference between TF-IDF and Okapi BM25 is obtained for all the queries as the $SF$ increases (Figure~\ref{fig:TFIDF_okapi_PostgreSQL_docs}). Moreover, the performance time for computing Okapi BM25 is smaller in all test scenarios than when computing TF-IDF. This is unexpected, as query complexity for computing the top-$k$ documents with TF-IDF is smaller then when using Okapi BM25. We hypothesize that such results are a direct consequence of how the DBMS builds query execution plans, with Oracle being the most proficient.

\paragraph{Database Implementation Comparison}

	This set of experiments presents a time performance comparison of database implementations w.r.t. $SF$ and weighting schemes for the top-$k$ keywords queries. We also test the performance of the distributed MongoDB environment, labeled \emph{Dist}. Moreover, MongoDB queries use both the AP and MR implementations for TF-IDF, but only the MR implementation for Okapi BM25. 

\begin{figure}[!htbp]
    \centering
    \begin{subfigure}{\columnwidth}
      \centering
        \includegraphics[width=\columnwidth]{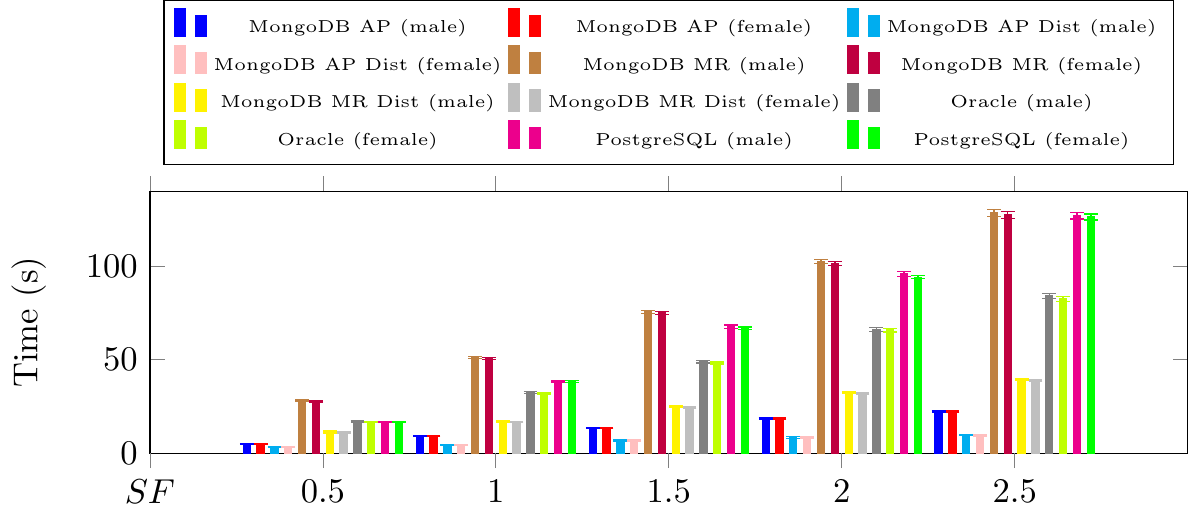}
        \caption{Q1 TF-IDF}
        \label{fig:TFIDF_q1}
    \end{subfigure}
    \begin{subfigure}{\columnwidth}
      \centering
        \includegraphics[width=\columnwidth]{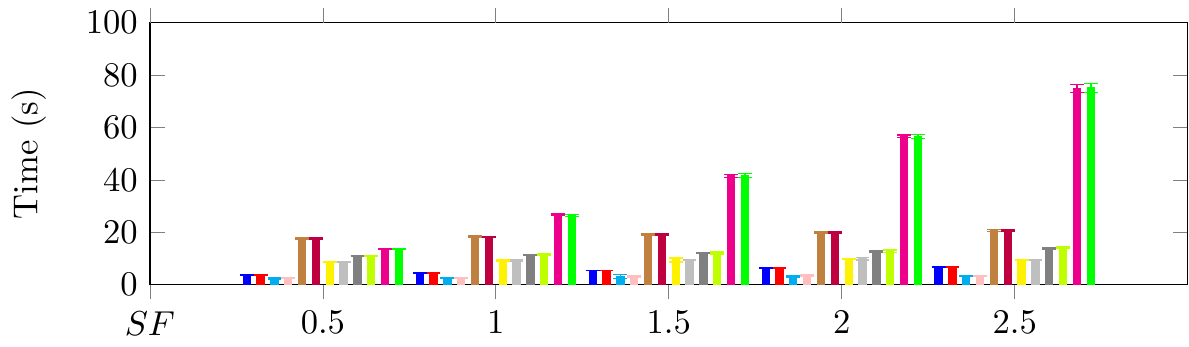}
        \caption{Q2 TF-IDF}
        \label{fig:TFIDF_q2}
    \end{subfigure}
    \begin{subfigure}{\columnwidth}
       \centering
        \includegraphics[width=\columnwidth]{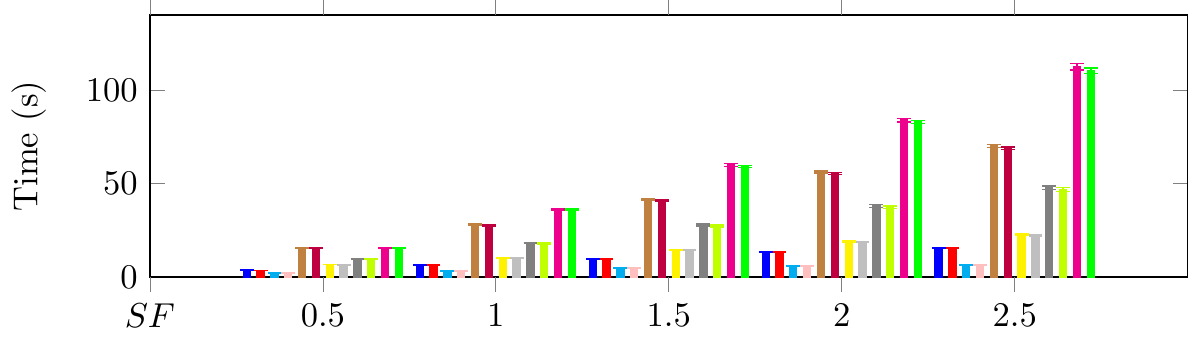}
        \caption{Q3 TF-IDF}
        \label{fig:TFIDF_q3}
    \end{subfigure}
    \begin{subfigure}{\columnwidth}
      \centering
        \includegraphics[width=\columnwidth]{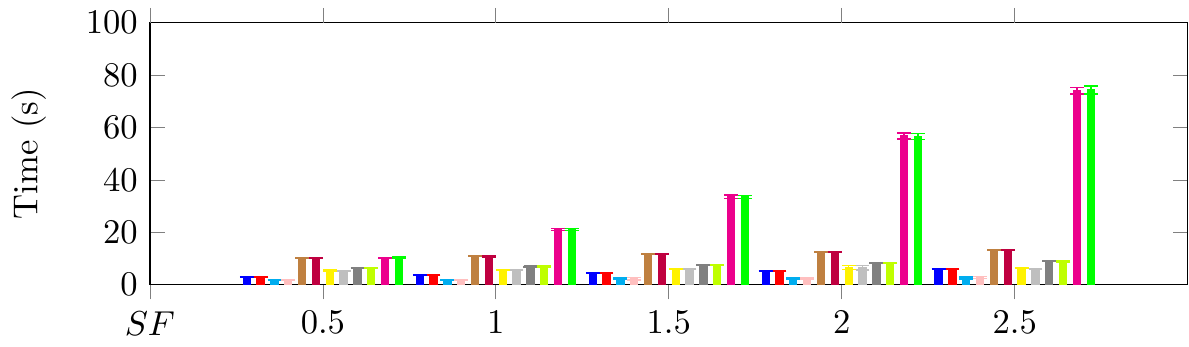}
        \caption{Q4 TF-IDF}
        \label{fig:TFIDF_q4}
    \end{subfigure}
    \caption{Top-$k$ keywords query comparison: TF-IDF}
    \label{fig:TFIDF_queries}
\end{figure}

\begin{figure}[!htbp]
    \centering
    \begin{subfigure}{\columnwidth}
      \centering
        \includegraphics[width=\columnwidth]{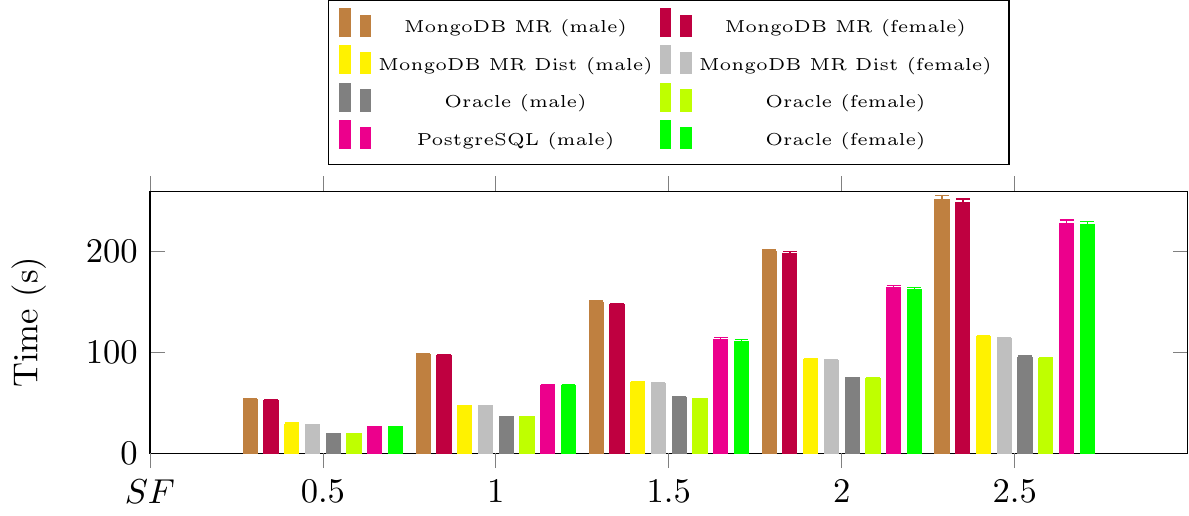}
        \caption{Q1 Okapi BM25}
        \label{fig:okapi_q1}
    \end{subfigure}
    \begin{subfigure}{\columnwidth}
      \centering
        \includegraphics[width=\columnwidth]{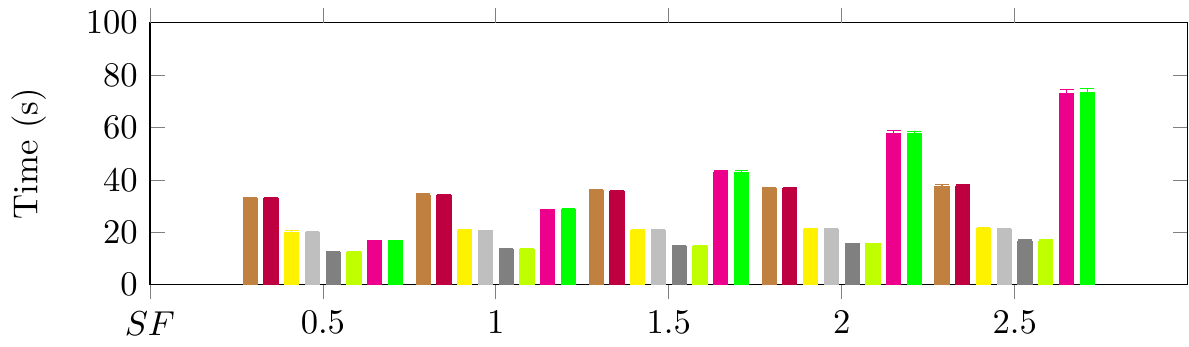}
        \caption{Q2 Okapi BM25}
        \label{fig:okapi_q2}
    \end{subfigure}
    \begin{subfigure}{\columnwidth}
       \centering
        \includegraphics[width=\columnwidth]{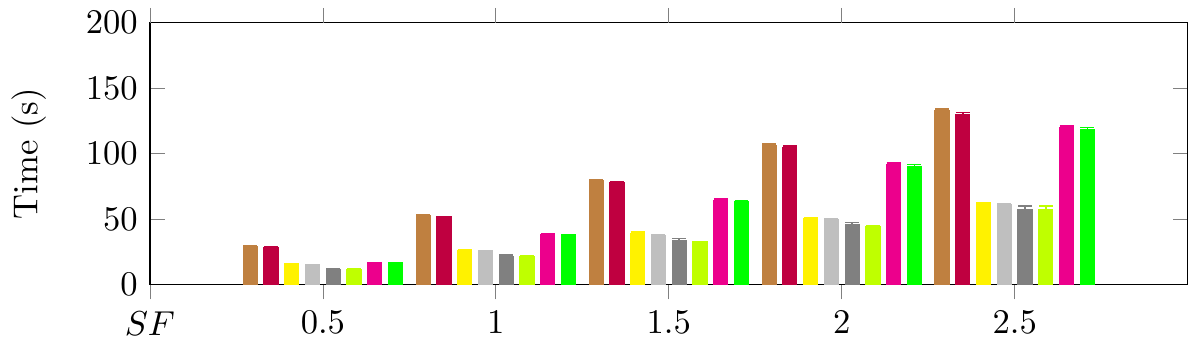}
        \caption{Q3 Okapi BM25}
        \label{fig:okapi_q3}
    \end{subfigure}
    \begin{subfigure}{\columnwidth}
      \centering
        \includegraphics[width=\columnwidth]{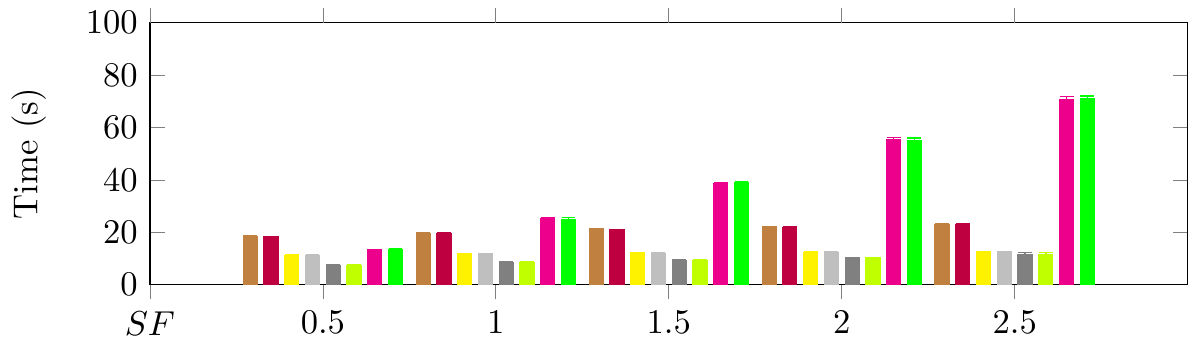}
        \caption{Q4 Okapi BM25}
        \label{fig:okapi_q4}
    \end{subfigure}
    \caption{Top-$k$ keywords query comparison: Okapi BM25}
    \label{fig:okapi_queries}
\end{figure}

    Oracle outperforms PostgreSQL in the single-instance environment for all test cases. Query execution time in Oracle is indeed half of the time than in PostgreSQL in the worse case (Figures~\ref{fig:TFIDF_queries} and~\ref{fig:okapi_queries}). Our hypothesis is that Oracle performs some optimization while PostgreSQL does not, but we could not verity it.

    In MongoDB, using AP instead of MR for computing TF-IDF  yields better execution time. In the worst case, AP execution time is indeed half of the time than MR's (Figure~\ref{fig:TFIDF_queries}), because MongoDB is optimized to use AP instead of MR. Thus, AP provides an alternative to MR and may be preferred sfor aggregation tasks, where the complexity of MP may be unwarranted. However, AP bears some limitations on value types, result size~\cite{MongoDBDocumentation} and memory use~\cite{Chodorow2013}.
    
    In the distributed environment, computing TF-IDF  using AP takes almost half the time than when using MR for all queries, too, regardless of $SF$ (Figure~\ref{fig:TFIDF_queries}), because AP is also optimized to perform better in a distributed environment than MR, though in both cases, the workload is distributed among all shards.

\begin{figure}[!htbp]
    \centering
    \begin{subfigure}{\columnwidth}
      \centering
        \includegraphics[width=\columnwidth]{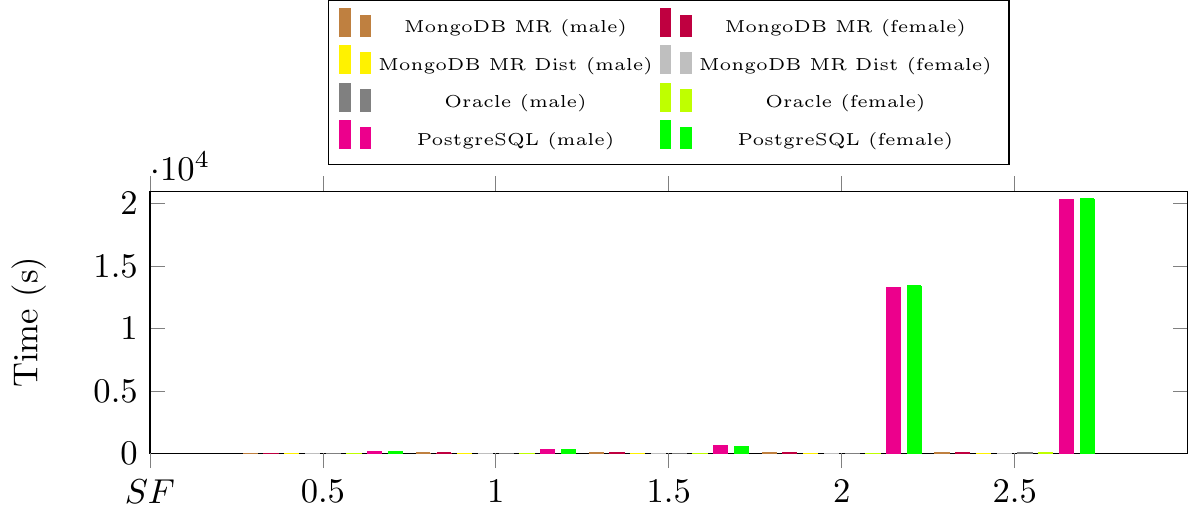}
        \caption{Q1 TF-IDF}
        \label{fig:TFIDF_q1_docs}
    \end{subfigure}
    \begin{subfigure}{\columnwidth}
      \centering
        \includegraphics[width=\columnwidth]{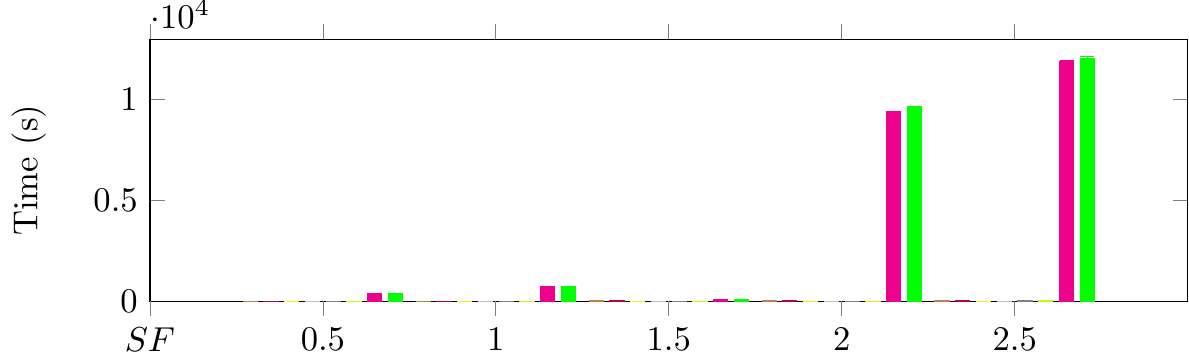}
        \caption{Q2 TF-IDF}
        \label{fig:TFIDF_q2_docs}
    \end{subfigure}
    \begin{subfigure}{\columnwidth}
       \centering
        \includegraphics[width=\columnwidth]{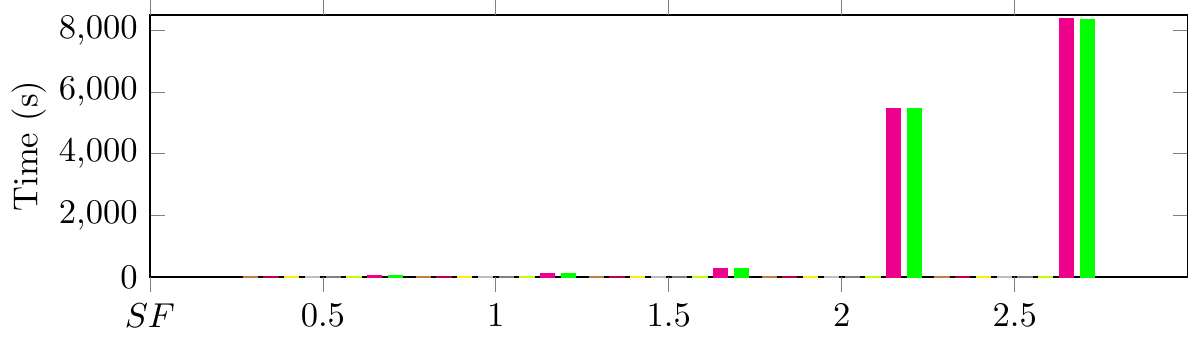}
        \caption{Q3 TF-IDF}
        \label{fig:TFIDF_q3_docs}
    \end{subfigure}
    \begin{subfigure}{\columnwidth}
      \centering
        \includegraphics[width=\columnwidth]{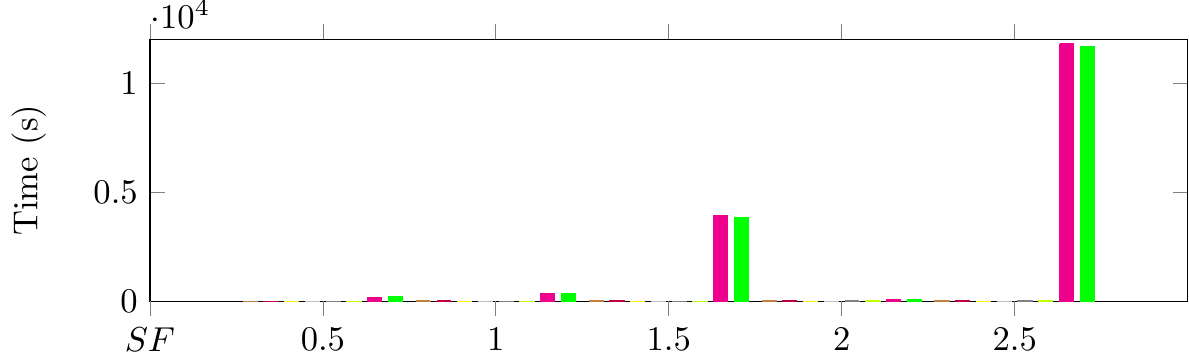}
        \caption{Q4 TF-IDF}
        \label{fig:TFIDF_q4_docs}
    \end{subfigure}
    \caption{Top-$k$ documents query comparison: TF-IDF}
    \label{fig:TFIDF_queries_docs}
\end{figure}

\begin{figure}[!htbp]
    \centering
    \begin{subfigure}{\columnwidth}
      \centering
        \includegraphics[width=\columnwidth]{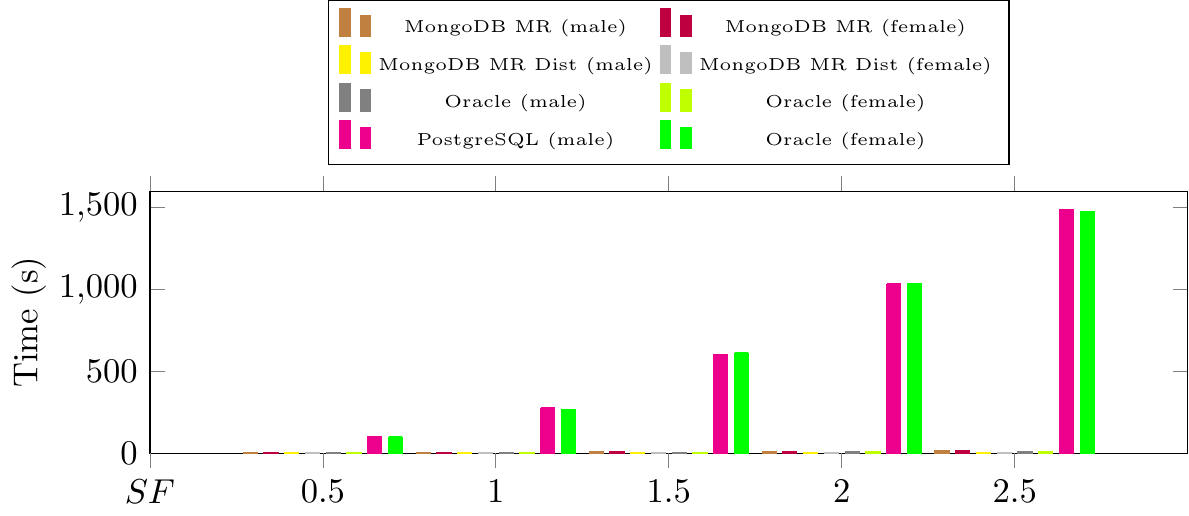}
        \caption{Q1 Okapi BM25}
        \label{fig:okapi_q1_docs}
    \end{subfigure}
    \begin{subfigure}{\columnwidth}
      \centering
        \includegraphics[width=\columnwidth]{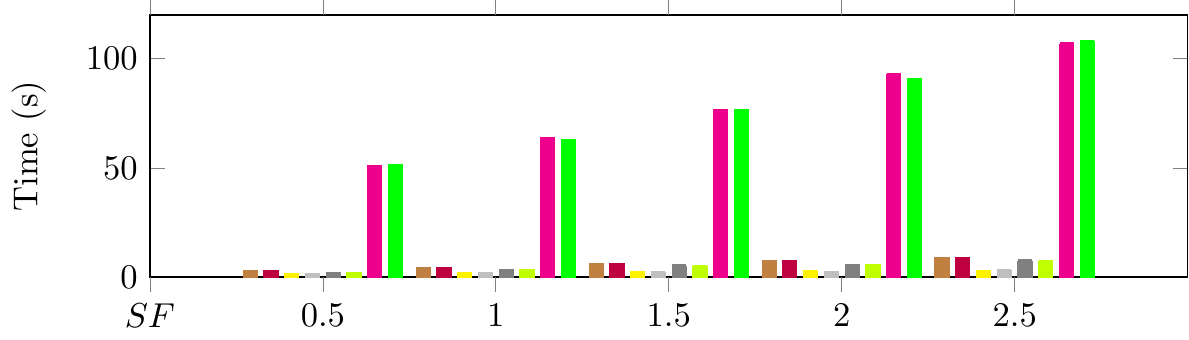}
        \caption{Q2 Okapi BM25}
        \label{fig:okapi_q2_docs}
    \end{subfigure}
    \begin{subfigure}{\columnwidth}
       \centering
        \includegraphics[width=\columnwidth]{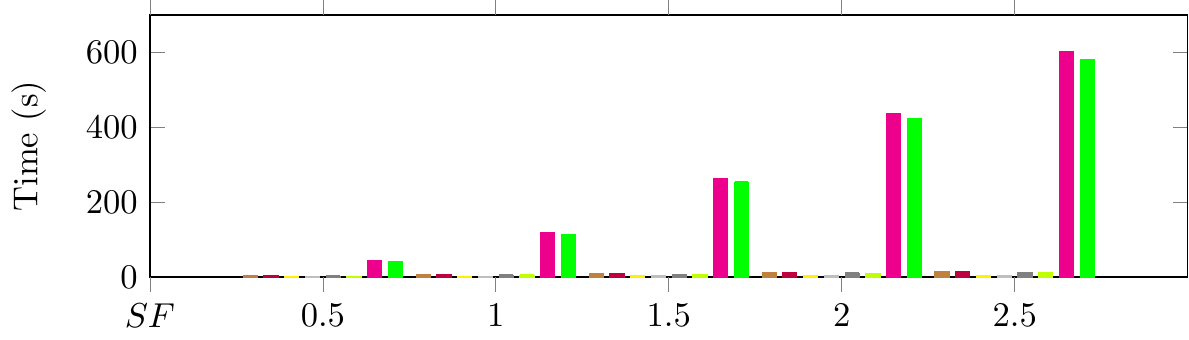}
        \caption{Q3 Okapi BM25}
        \label{fig:okapi_q3_docs}
    \end{subfigure}
    \begin{subfigure}{\columnwidth}
      \centering
        \includegraphics[width=\columnwidth]{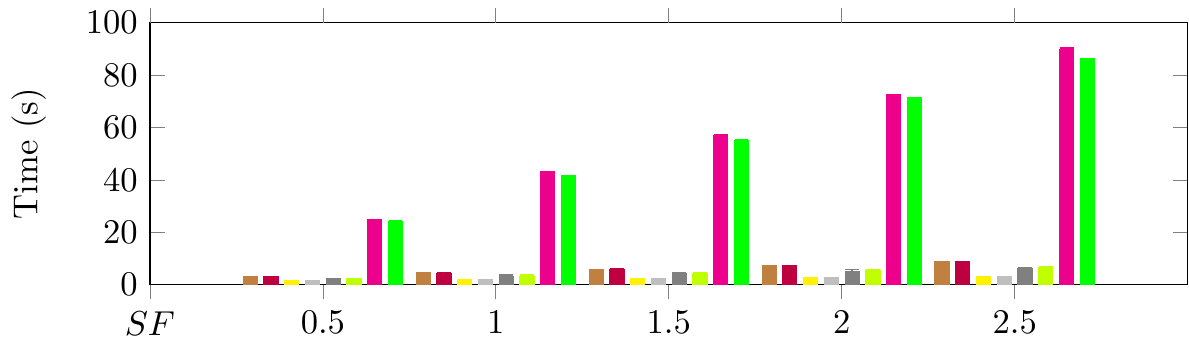}
        \caption{Q4 Okapi BM25}
        \label{fig:okapi_q4_docs}
    \end{subfigure}
    \caption{Top-$k$ documents query comparison: Okapi BM25}
    \label{fig:okapi_queries_docs}
\end{figure}

	Figure~\ref{fig:TFIDF_queries_docs} presents a time performance comparison of database implementations w.r.t. $SF$ and weighting schemes for the top-$k$ documents queries that use the TF-IDF weighting schema, while Figure~\ref{fig:okapi_queries_docs} for the queries that use the Okapi BM25 weighting schema.

	PostgreSQL has the worst execution time regardless of the weighting schema or $SF$. Our hypothesis is that PostgreSQL does not optimize the query execution plan due to the absence of indexes.
	
	Oracle outperforms MongoDB in a single instance environment for both weighting schemas. Moreover, Oracle is optimized to make the best use of  system resources in a single instance environment~\cite{Bhatiya2017}.
	
	MongoDB in a distributed environment has the best performance time regardless of the weighting schema. Query execution performance in MongoDB is improved because the workload is distributed between shards and results are aggregated after each shard finishes a task.

\subsubsection{T$^2$K$^2$D$^2$}

	\paragraph{Weighting schema comparison} 
	
\begin{figure}[!htbp]
    \centering
    \begin{subfigure}{\columnwidth}
      \centering
        \includegraphics[width=\columnwidth]{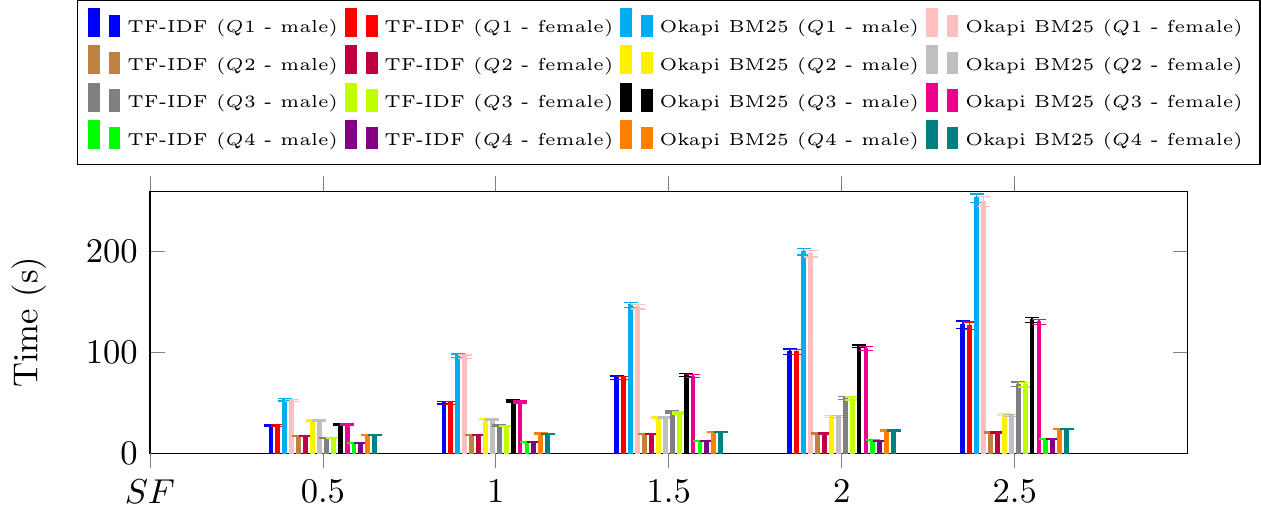}
        \caption{MongoDB MR OLAP: TF-IDF vs. Okapi BM25}
        \label{fig:TFIDF_okapi_MR_OLAP}     
    \end{subfigure}
    \begin{subfigure}{\columnwidth}
      \centering
        \includegraphics[width=\columnwidth]{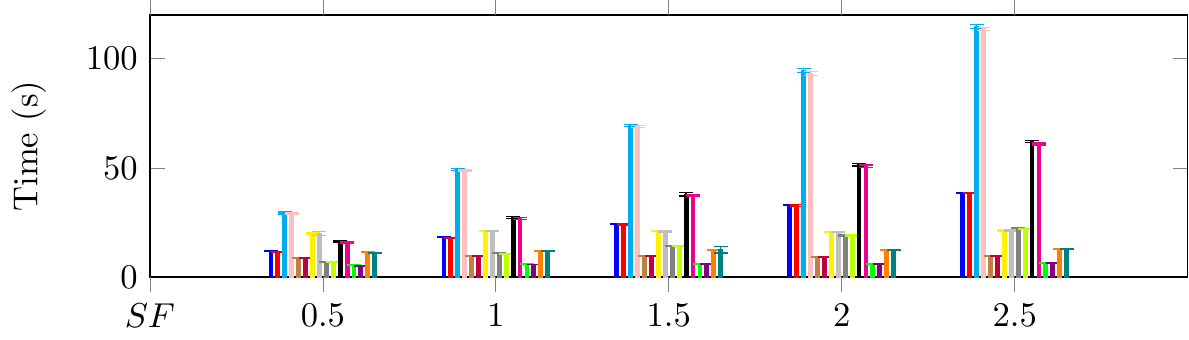}
        \caption{MongoDB MR Distributed OLAP: TF-IDF vs. Okapi BM25}
        \label{fig:TFIDF_okapi_MR_dist_OLAP}
    \end{subfigure}
    \begin{subfigure}{\columnwidth}
      \centering
        \includegraphics[width=\columnwidth]{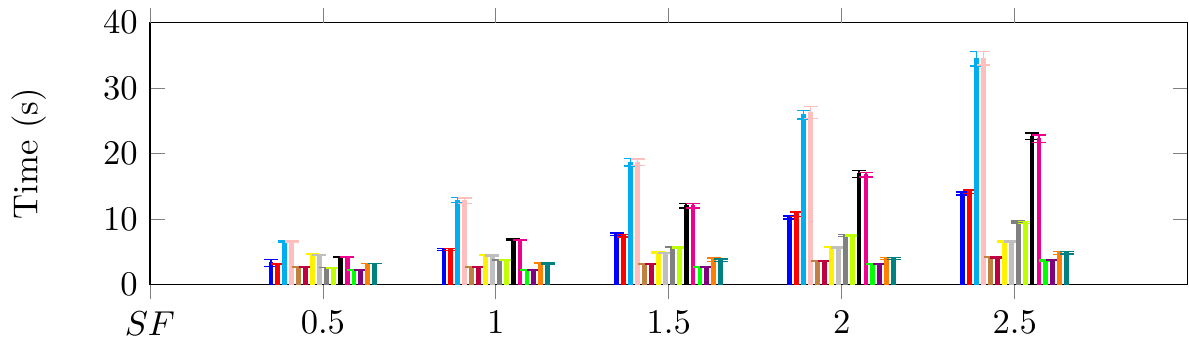}
        \caption{Oracle OLAP: TF-IDF vs. Okapi BM25}
        \label{fig:TFIDF_okapi_Oracle_OLAP}
    \end{subfigure}    
    \begin{subfigure}{\columnwidth}
      \centering
				\includegraphics[width=\columnwidth]{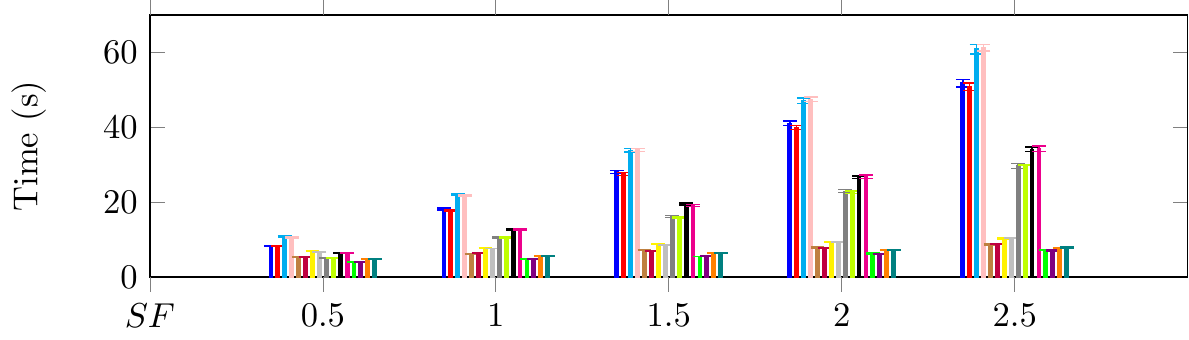}
            \caption{PostgreSQL OLAP: TF-IDF vs. Okapi BM25}
            \label{fig:TFIDF_okapi_PostgreSQL_OLAP}
    \end{subfigure}
    \caption{OLAP top-$k$ keywords TF-IDF vs. Okapi BM25 comparison}
    \label{fig:TFIDF_okapi_olap}
\end{figure}

\begin{figure}[!htbp]
    \centering
    \begin{subfigure}{\columnwidth}
      \centering
        \includegraphics[width=\columnwidth]{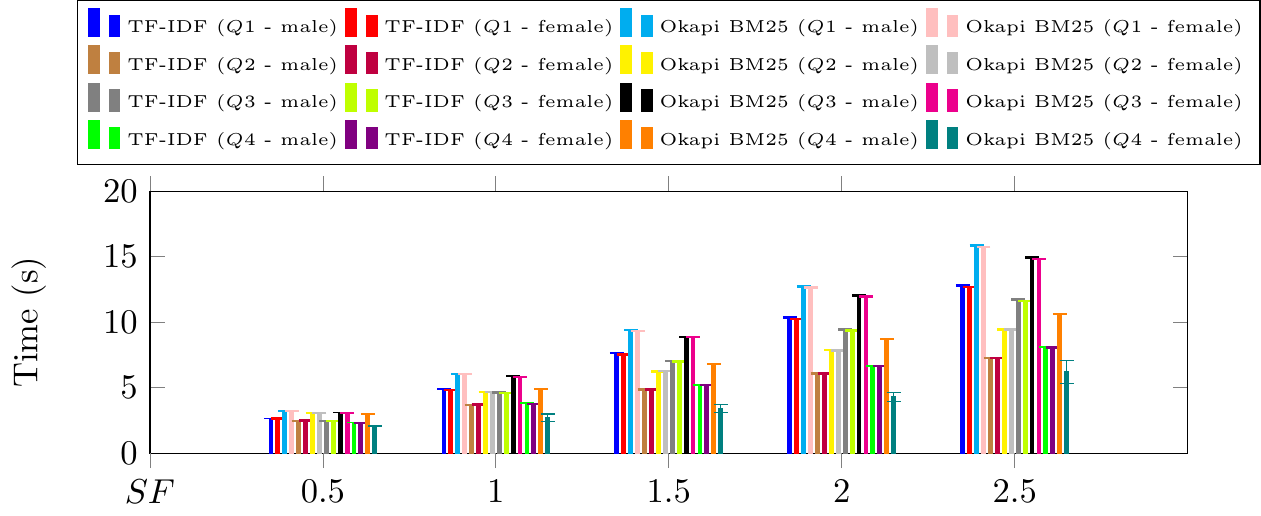}\textbf{}
        \caption{MongoDB MR OLAP: TF-IDF vs. Okapi BM25}
        \label{fig:TFIDF_okapi_MR_OLAP_docs}     
    \end{subfigure}
    \begin{subfigure}{\columnwidth}
      \centering
        \includegraphics[width=\columnwidth]{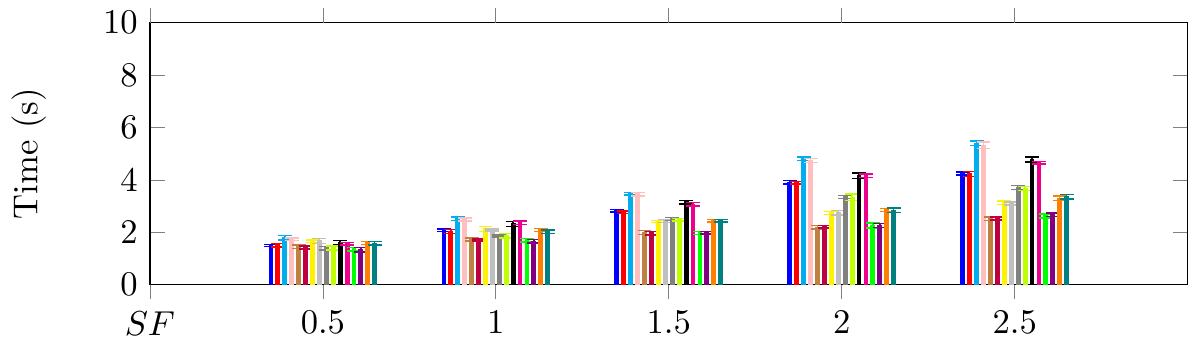}
        \caption{MongoDB MR Distributed OLAP: TF-IDF vs. Okapi BM25}
        \label{fig:TFIDF_okapi_MR_dist_OLAP_docs}
    \end{subfigure}
    \begin{subfigure}{\columnwidth}
      \centering
        \includegraphics[width=\columnwidth]{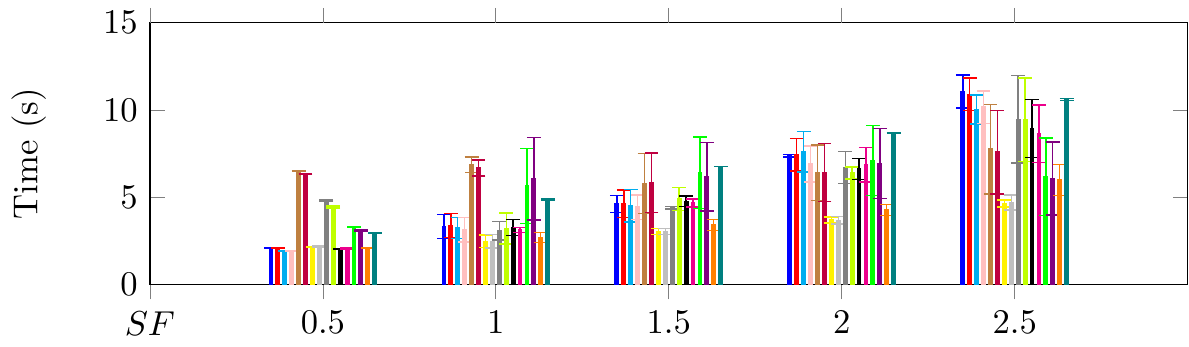}
        \caption{Oracle OLAP: TF-IDF vs. Okapi BM25}
        \label{fig:TFIDF_okapi_Oracle_OLAP_docs}
    \end{subfigure}    
    \begin{subfigure}{\columnwidth}
      \centering
				\includegraphics[width=\columnwidth]{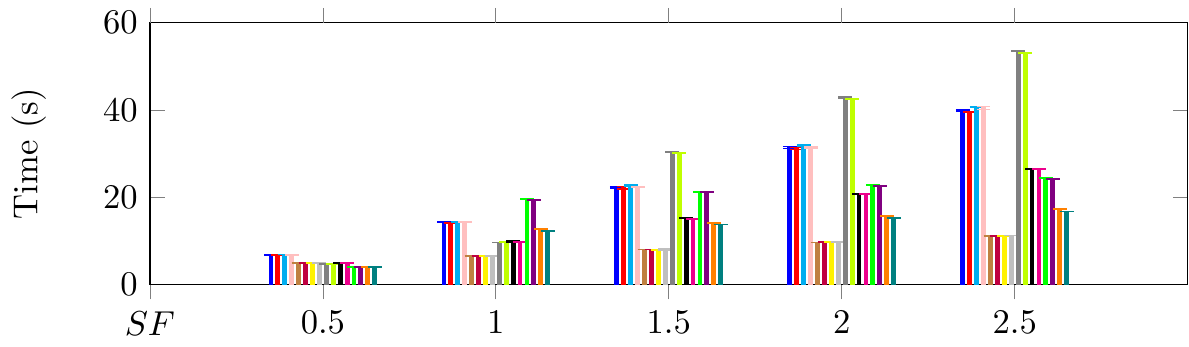}
            \caption{PostgreSQL OLAP: TF-IDF vs. Okapi BM25}
            \label{fig:TFIDF_okapi_PostgreSQL_OLAP_docs}
    \end{subfigure}
    \caption{OLAP top-$k$ documents TF-IDF vs. Okapi BM25 comparison}
    \label{fig:TFIDF_okapi_olap_docs}
\end{figure}

	Figure~\ref{fig:TFIDF_okapi_olap} present a comparison by database system and weighting schema of query response time for retrieving the top-$k$ keywords w.r.t. scale factor $SF$ for the multidimensional implementation, while Figure~\ref{fig:TFIDF_okapi_olap_docs} for retrieving top-$k$ documents. 
	
	For MongoDB the same patterns in performance appear as for the T$^2$K$^2$ test scenarios. The time performance is almost non-existent between the queries run on T$^2$K$^2$ (Figures~\ref{fig:TFIDF_okapi_MR} and~\ref{fig:TFIDF_okapi_MR_dist}) and the ones executed on T$^2$K$^2$D$^2$ (Figures~\ref{fig:TFIDF_okapi_MR_OLAP} and~\ref{fig:TFIDF_okapi_MR_dist_OLAP}). This pattern emerges because MongoDB is a schemaless database and the data modeling does not really influence performance. The same results are optained for the top-$k$ documents (Figures~\ref{fig:TFIDF_okapi_MR_OLAP_docs} and~\ref{fig:TFIDF_okapi_MR_dist_OLAP_docs}).
	
	Using a multidimensional model improves performance with both Oracle and PostgreSQL. Using such a model, the indexes used for primary and foreign keys to associate entities make query execution plan optimization easy, because a join does not have to pass through multiple tables as in the normalized model.
    
	For Oracle, the execution time of queries $Q1$ and $Q3$ that use TF-IDF is half the time of the execution time of the same queries that use Okapi BM25 for top-$k$ keywords (Figure~\ref{fig:TFIDF_okapi_Oracle_OLAP}) w.r.t. $SF$. For top-$k$ documents, the execution time for queries $Q1$ and $Q2$ is almost the same regardless of weight. It should be noted that Oracle's query execution time for top-$k$ documents is unstable, as the standard deviation fluctuates. We think this fluctuation might be a direct consequence of cache and buffer cleaning.

	PostgreSQL query execution time is also improved when using a multidimensional model. The query execution time are stable with a low standard deviation. For top-$k$ documents, $Q1$ has the worst execution time. Moreover, the execution time between the queries that compute the weights with TF-IDF and the ones that compute the weights with Okapi BM25 has a factor of $\sim$1.2 regardless of $SF$. For top-$k$ documents, $Q1$ execution time is the same regardless of weighting schema. The execution time of $Q3$ with Okapi BM25 is half the time of the execution of the same query with TF-IDF.

    \paragraph{Database implementation comparison.}
    
\begin{figure}[!htbp]
    \centering
    \begin{subfigure}{\columnwidth}
      \centering
        \includegraphics[width=\columnwidth]{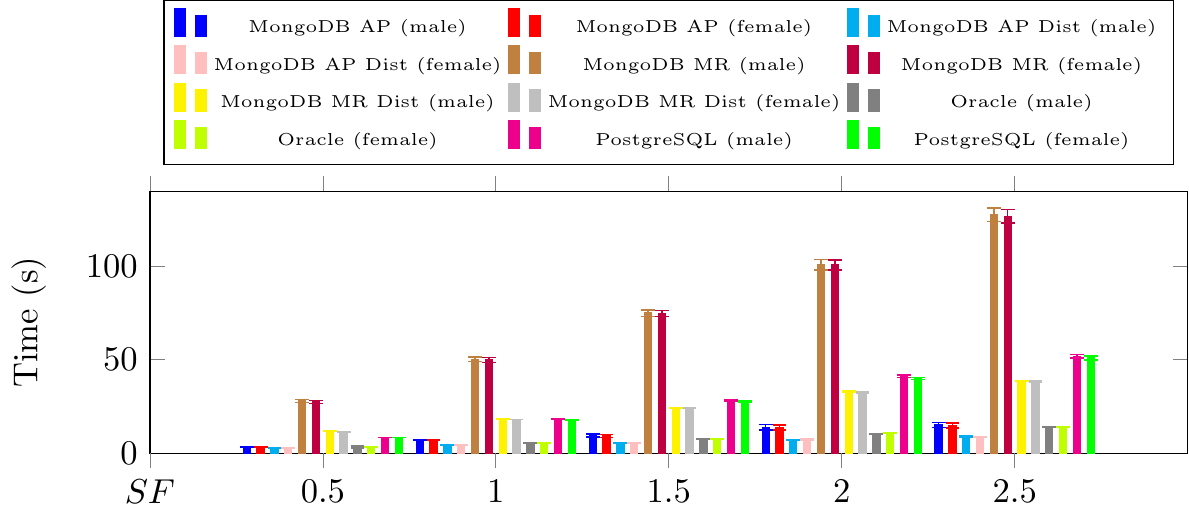}
        \caption{Q1 TF-IDF OLAP}
        \label{fig:TFIDF_q1_OLAP}
    \end{subfigure}
    \begin{subfigure}{\columnwidth}
      \centering
        \includegraphics[width=\columnwidth]{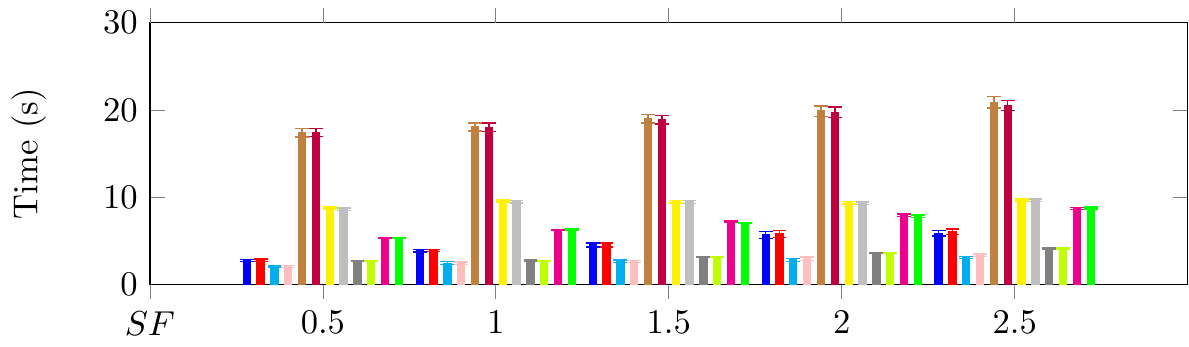}
        \caption{Q2 TF-IDF OLAP}
        \label{fig:TFIDF_q2_OLAP}
    \end{subfigure}
    \begin{subfigure}{\columnwidth}
       \centering
        \includegraphics[width=\columnwidth]{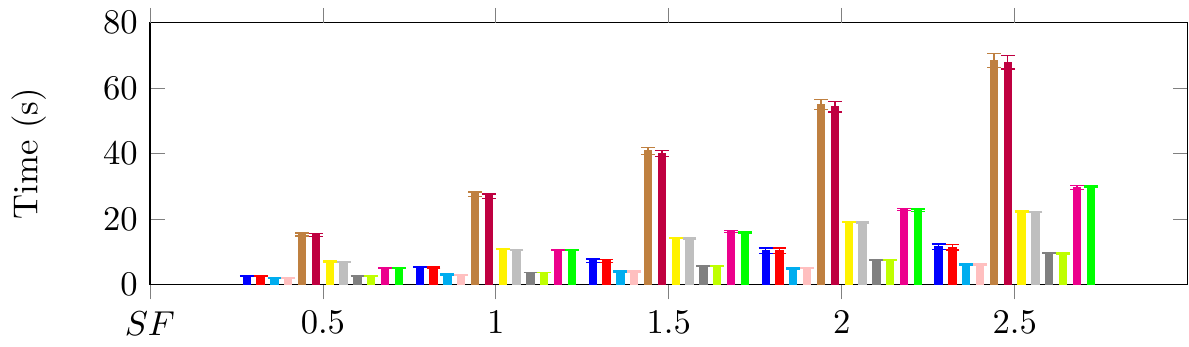}
        \caption{Q3 TF-IDF OLAP}
        \label{fig:TFIDF_q3_OLAP}
    \end{subfigure}
    \begin{subfigure}{\columnwidth}
      \centering
        \includegraphics[width=\columnwidth]{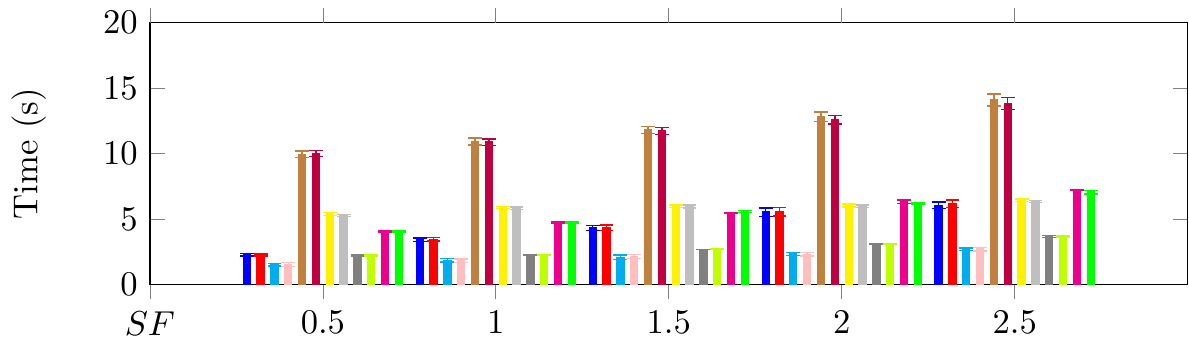}
        \caption{Q4 TF-IDF OLAP}
        \label{fig:TFIDF_q4_OLAP}
    \end{subfigure}
    \caption{Top-$k$ keywords OLAP query comparison: TF-IDF}
    \label{fig:TFIDF_queries_olap}
\end{figure}

\begin{figure}[!htbp]
    \centering
    \begin{subfigure}{\columnwidth}
      \centering
        \includegraphics[width=\columnwidth]{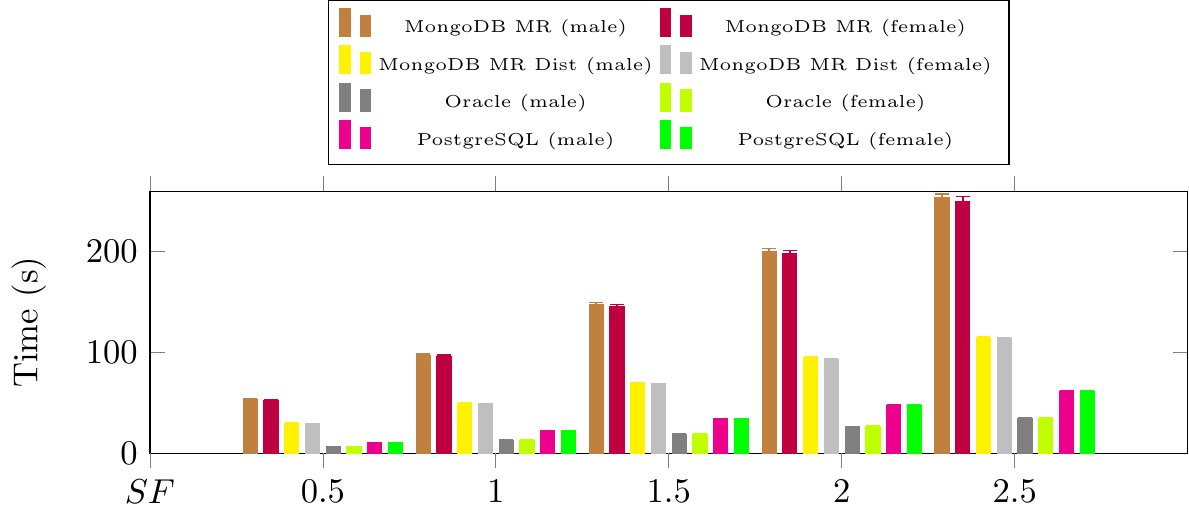}
        \caption{Q1 Okapi BM25 OLAP}
        \label{fig:okapi_q1_OLAP}
    \end{subfigure}
    \begin{subfigure}{\columnwidth}
      \centering
        \includegraphics[width=\columnwidth]{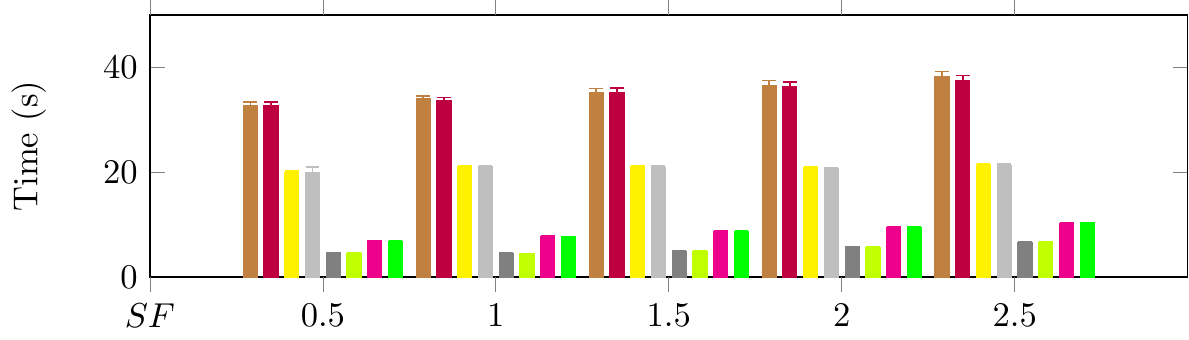}
        \caption{Q2 Okapi BM25 OLAP}
        \label{fig:okapi_q2_OLAP}
    \end{subfigure}
    \begin{subfigure}{\columnwidth}
       \centering
        \includegraphics[width=\columnwidth]{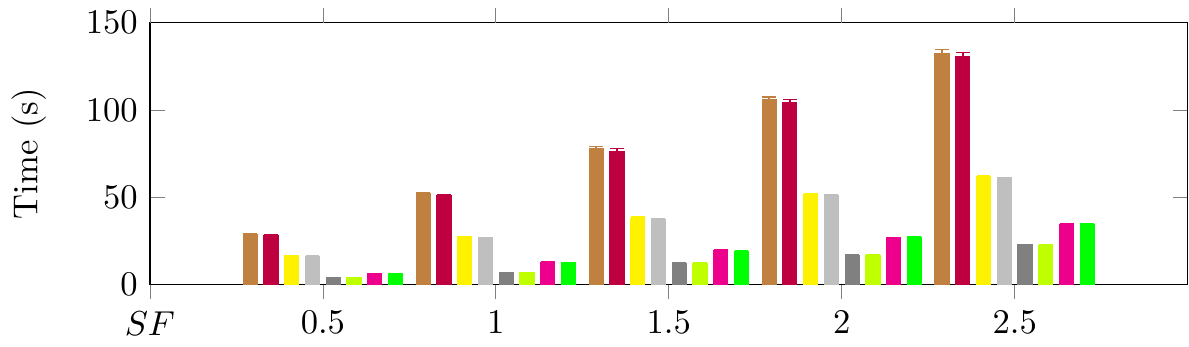}
        \caption{Q3 Okapi BM 25 OLAP}
        \label{fig:okapi_q3_OLAP}
    \end{subfigure}
    \begin{subfigure}{\columnwidth}
      \centering
        \includegraphics[width=\columnwidth]{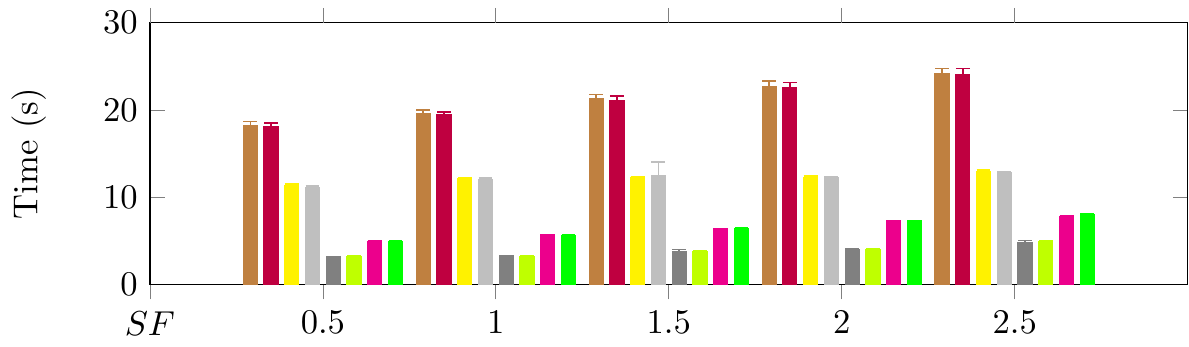}
        \caption{Q4 Okapi BM25 OLAP}
        \label{fig:okapi_q4_OLAP}
    \end{subfigure}
    \caption{Top-$k$ keywords OLAP query comparison: Okapi BM25}
    \label{fig:okapi_queries_olap}
\end{figure}

\begin{figure}[!htbp]
    \centering
    \begin{subfigure}{\columnwidth}
      \centering
        \includegraphics[width=\columnwidth]{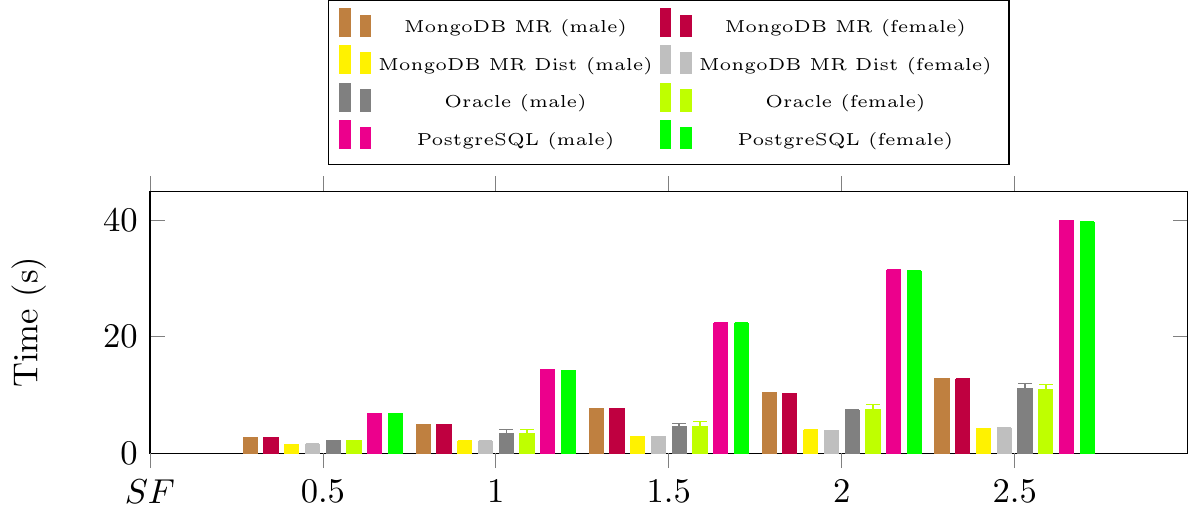}
        \caption{Q1 TF-IDF OLAP}
        \label{fig:TFIDF_q1_OLAP_docs}
    \end{subfigure}
    \begin{subfigure}{\columnwidth}
      \centering
        \textbf{\includegraphics[width=\columnwidth]{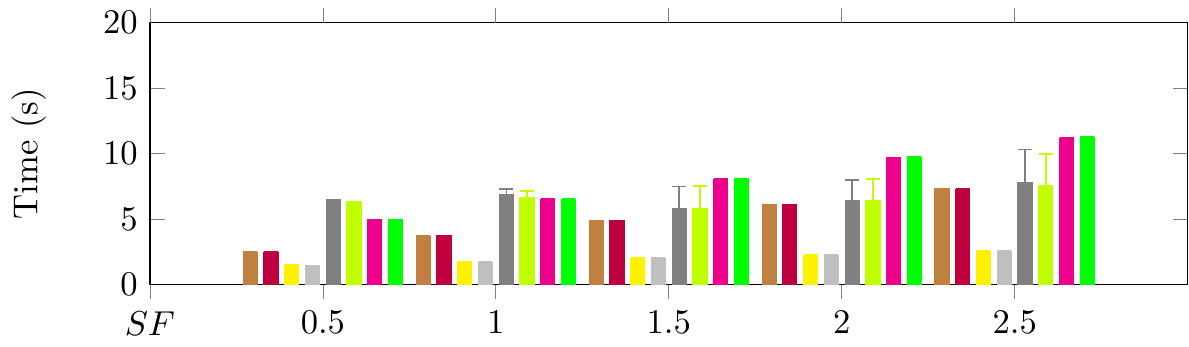}}
        \caption{Q2 TF-IDF OLAP}
        \label{fig:TFIDF_q2_OLAP_docs}
    \end{subfigure}
    \begin{subfigure}{\columnwidth}
       \centering
        \includegraphics[width=\columnwidth]{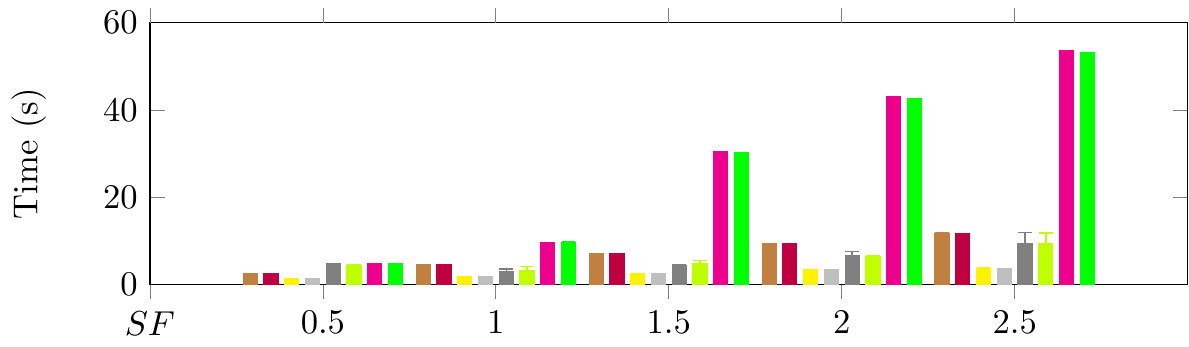}
        \caption{Q3 TF-IDF OLAP}
        \label{fig:TFIDF_q3_OLAP_docs}
    \end{subfigure}
    \begin{subfigure}{\columnwidth}
      \centering
        \includegraphics[width=\columnwidth]{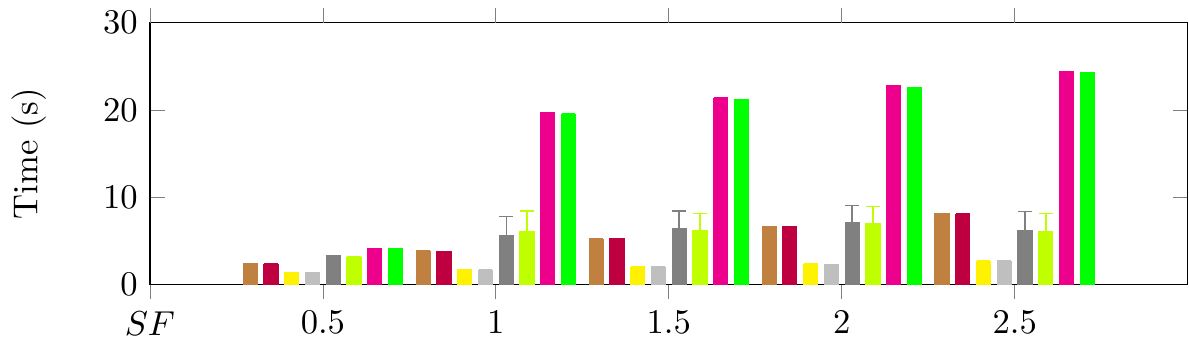}
        \caption{Q4 TF-IDF OLAP}
        \label{fig:TFIDF_q4_OLAP_docs}
    \end{subfigure}
    \caption{Top-$k$ documents OLAP query comparison: TF-IDF}
    \label{fig:TFIDF_queries_olap_docs}
\end{figure}

	Oracle has the best execution time regardless of weighting schema in a single instance for both top-$k$ keywords and top-$k$ documents (Figures~\ref{fig:TFIDF_queries_olap}, \ref{fig:okapi_queries_olap}, \ref{fig:TFIDF_queries_olap_docs} and~\ref{fig:okapi_queries_olap}). Besides the resource allocation optimization used by Oracle in a single instance environment, query complexity also plays an important role in query execution performance, as the number of joins decreases when using a multidimensional model.
		
	MongoDB in a distributed environment has the overall best performance time for the queries that use the TF-IDF weighting schema, but it is outperformed by both Oracle and PostgreSQL when using Okapi BM25. Because MongoDB uses a schemaless flexible model, the design and implementation of the schema indeed do not really affect query performance results. Furthermore, query complexity, which depends a lot on the weighting scheme used (Tables~\ref{tbl:query_complexity_star} and~\ref{tbl:query_complexity_star_docs}), plays an important role. In Oracle and PostgreSQL, Okapi BM25 weights are computed through nested queries, but in MongoDB, separate queries are needed. To improve MongoDB query performance, an indexing and sharding key analysis should be made, but such a extensive study is outside the scope of this paper.
	
	MongoDB in a single environment has the overall worst performance between the tested databases for top-$k$ keywords (Figures~\ref{fig:TFIDF_queries_olap} and~\ref{fig:okapi_queries_olap}), while PostgreSQL for top-$k$ documents (Figures~\ref{fig:TFIDF_queries_olap_docs} and~\ref{fig:okapi_queries_olap}). 
	
	Using a multidimensional model significantly improves query performance in PostgreSQL, because query complexity is decreased. The multidimensional model used in T$^2$K$^2$D$^2$ indeed only features one-to-many relationships at the conceptual level, and thus does not include any bridge tables at the logical level for expressing many-to-many relationships, as in T$^2$K$^2$. Therefore, the number of joins in queries decreases and so does query complexity.

\begin{figure}[!htbp]
    \centering
    \begin{subfigure}{\columnwidth}
      \centering
        \includegraphics[width=\columnwidth]{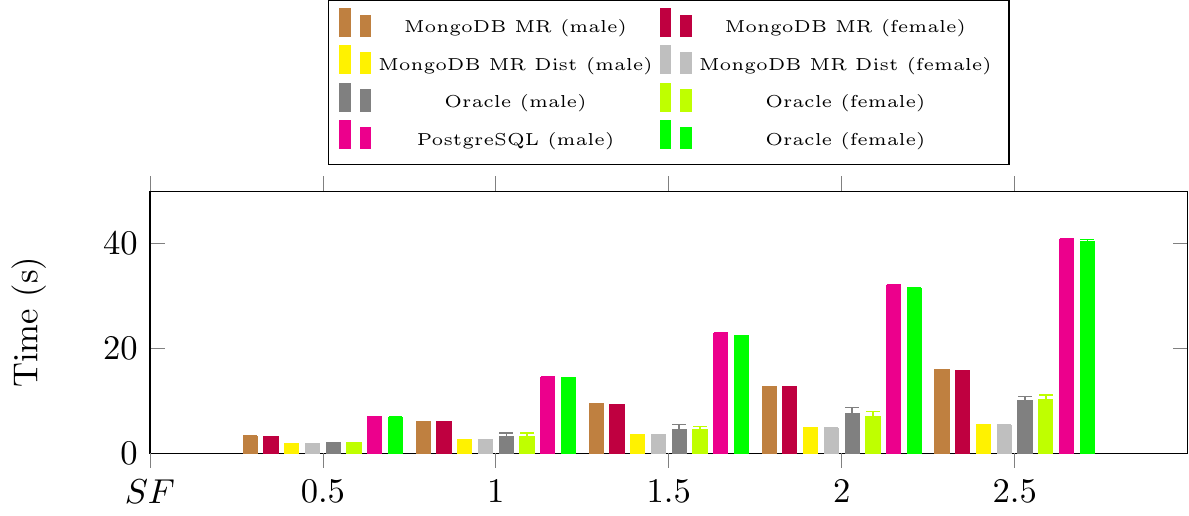}
        \caption{Q1 Okapi BM25 OLAP}
        \label{fig:okapi_q1_OLAP_docs}
    \end{subfigure}
    \begin{subfigure}{\columnwidth}
      \centering
        \includegraphics[width=\columnwidth]{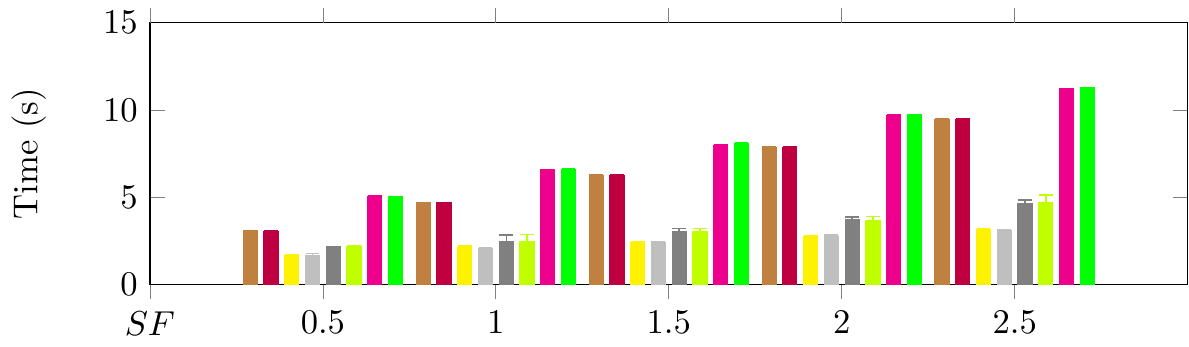}
        \caption{Q2 Okapi BM25 OLAP}
        \label{fig:okapi_q2_OLAP_docs}
    \end{subfigure}
    \begin{subfigure}{\columnwidth}
       \centering
        \includegraphics[width=\columnwidth]{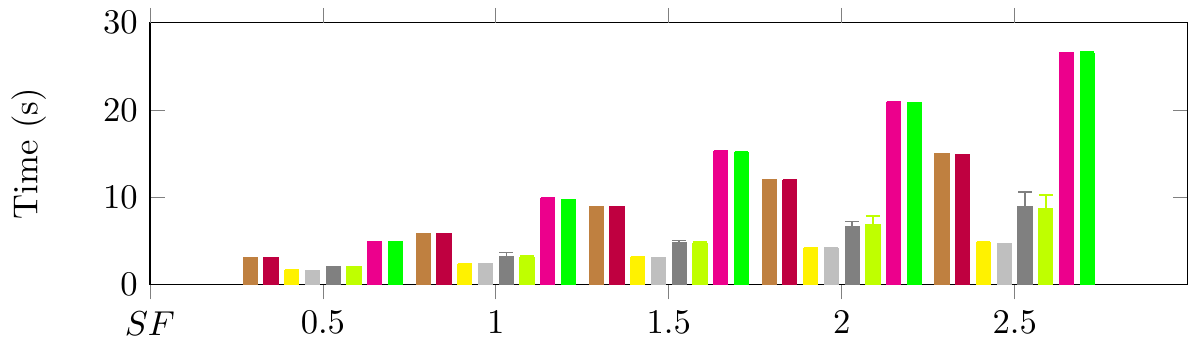}
        \caption{Q3 Okapi BM 25 OLAP}
        \label{fig:okapi_q3_OLAP_docs}
    \end{subfigure}
    \begin{subfigure}{\columnwidth}
      \centering
        \includegraphics[width=\columnwidth]{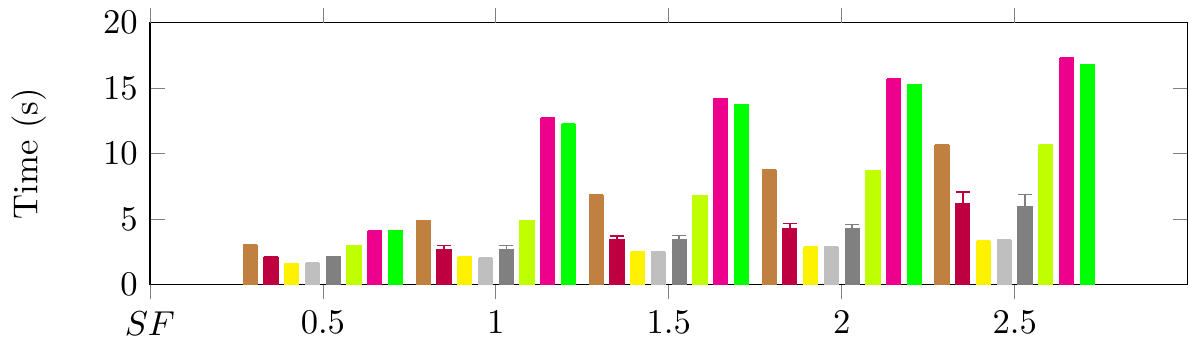}
        \caption{Q4 Okapi BM25 OLAP}
        \label{fig:okapi_q4_OLAP_docs}
    \end{subfigure}
    \caption{Top-$k$ documents OLAP query comparison: Okapi BM25}
    \label{fig:okapi_queries_olap_docs}
\end{figure}

\section{Conclusion}
\label{sec:Conclusion}

	Jim Gray defined four primary criteria to specify a ``good'' benchmark \cite{Gray1993}:
	\begin{itemize}
		\item \textit{Relevance:} The benchmark must deal with aspects of performance that appeal to the largest number of users. Considering the wide usage of top-$k$ keywords and documents queries in various text analytics tasks, we think T$^2$K$^2$ and T$^2$K$^2$D$^2$ fulfills this criterion. We also show in Section~\ref{sec:Experiments} that our benchmark achieves what it is designed for.
		\item \textit{Portability:} The benchmark must be reusable to test the performances of different database systems. We successfully used T$^2$K$^2$ and T$^2$K$^2$D$^2$ to compare two types of database systems, namely relational and document-oriented systems.
		\item \textit{Simplicity:} The benchmark must be feasible and must not require too many resources. We designed T$^2$K$^2$ and T$^2$K$^2$D$^2$ with this criterion in mind (Section~\ref{sec:Specs}), which is particularly important for reproducibility. We notably made up parameters that are easy to setup.
		\item \textit{Scalability:} The benchmark must adapt to small or large computer architectures. By introducing scale factor $SF$, we allow users to simply parameterize T$^2$K$^2$ and T$^2$K$^2$D$^2$ and achieve some scaling, though it could be pushed further in terms of data volume. 
	\end{itemize}

	Regarding experimental results, for T$^2$K$^2$'s top-$k$ keywords the best TF-IDF computing time in single-instance mode is obtained by MongoDB using PA. Oracle registers the best execution time among  relational databases. Moreover, the best solution for computing TF-IDF is to use MongoDB in a distributed environment because increasing the number of nodes expectingly helps decreasing query execution time. Oracle outperforms the other database systems when computing Okapi BM25. Moreover, MongoDB in single-instance mode has the worst execution time for computing Okapi BM25 with queries $Q1$ and $Q3$, while PostgreSQL has the worst execution time for queries $Q2$ and $Q4$. This is unexpected, because the trends should have remained the same for these two database systems. When computing the top-$k$ documents with T$^2$K$^2$, the best performance time is obtained by MongoDB in a distributed environment and by Oracle in a single instance regardless of the schema, while PostgreSQL has the worst performance. 
	
	When using a multidimensional model with T$^2$K$^2$D$^2$, MongoDB registers the same performance or both top-$k$ keywords and top-$k$ documents as with T$^2$K$^2$. Oracle in a single instance has the overall best performance for both tasks. For top-$k$ keywords MongoDB in a single instance has the worst performance while for top-$k$ documents, PostgreSQL has the worst performance. 
	
	In future work, we plan to expand T$^2$K$^2$ and T$^2$K$^2$D$^2$ dataset significantly to aim at big data-scale volume (scalability). We also plan to adapt our benchmarks so that they run in the Hadoop and Spark environments. Moreover, we intend to further our proof of concept and validation efforts by benchmarking other NoSQL database systems and gain insight regarding their capabilities and shortcomings (relevance). Eventually, we considered in this paper that TF-IDF and Okapi BM25 were enough for machine learning tasks. Yet, the next version of our benchmarks should include other weighting schemes, such as KL-divergence~\cite{Raiber2017}, also to improve their relevance.

\section*{References}
\bibliographystyle{elsarticle-num}
\bibliography{benchmark}

\end{document}